\documentclass[%
 reprint,
 amsmath,amssymb,
 aps,
]{revtex4-2}

\usepackage{graphicx}
\usepackage{float}
\usepackage[export]{adjustbox} 
\usepackage{subcaption}
\usepackage{subfig}
\usepackage{subfiles}
\usepackage{braket}
\usepackage{svg}
\usepackage{tikz}
    \usetikzlibrary{quantikz}

\usepackage{xcolor}
\usepackage{algorithm}
\usepackage{algpseudocode}
\usepackage{dcolumn}
\usepackage{bm}

\begin{document}

\preprint{APS/123-QED}

\title{Float Lattice Gas Automata: A connection between Molecular Dynamics and Lattice Boltzmann Method for quantum computers }

\author{Antonio David Bastida Zamora}
 \email{antonio.bastidazamora@quanscient.com}
 \altaffiliation[Also at ]{Aix-Marseille University}
\author{Ljubomir Budinski}%
  \altaffiliation[Also at ]{Faculty of Technical Sciences, University of Novi Sad}
  
\affiliation{Quanscient Oy,Finland}%

\author{Pierre Sagaut}%
\affiliation{Aix Marseille Univ, Centrale Med, CNRS, M2P2 Laboratory, 13013 Marseille, France}
  \altaffiliation[Also at ]{Quanscient Oy,Finland}
  
\author{Valtteri Lahtinen}
\affiliation{Quanscient Oy,Finland}

\date{\today}

\begin{abstract}

Building upon the Integer Lattice Gas Automata framework of Blommel \textit{et al.} \cite{PhysRevE.97.023310}, we introduce a simplified, fluctuation-free variant. This approach relies on floating-point numbers and closely mirrors the Lattice Boltzmann Method (LBM), with the key distinction being a novel collision operator. This operator, derived from the ensemble average of transition probabilities, generates nonlinear terms. We propose this new Float Lattice Gas Automata (FLGA) collision as a computationally efficient alternative to traditional and quantum LBM implementations.

\end{abstract}

\maketitle


\section{Introduction}
In the last part of the 20th century, Lattice Gas Automata 
(LGA)\footnote{We will use the acronym LGA for both the singular ''Lattice Gas Automaton´´ and the plural ''Lattice Gas Automata´´ as it is always clear from the context which form is intended.} were conceived as promising methods for computational fluid dynamics (CFD) due to their ability to model Navier-Stokes equations at a microscopic scale using simple and deterministic rules consisting of two steps: collision and propagation \cite{Hardy_Pomeau_1972}. It was soon when the scientific community realised that these methods suffer several problems, mainly the statistical noise, lack of Galilean invariance, and the pressure as a spurious function of the density and the velocity as an extra term. That, together with the computational cost of the model, made researchers focus on solving the problem from another perspective. Noticing that LGA are connected with the Boltzmann equation at a mesoscopic scale, which can be derived as a result of an ensemble average by performing a Chapman-Enskog expansion \cite{SUCCI1991219}, the idea of directly modelling the Boltzmann equation surged. As a result of this new research lead, the Lattice Boltzmann Method (LBM) arised as a successful numerical tool for CFD applications and other computational physics domains such as electromagnetism \cite{PhysRevE.82.056708}. Later, further developments such as the Bhatnagar-Gross-Krook (BGK) approximation \cite{Qian1992} or one-step simplified LBM \cite{shenglei2023novel} improved its efficiency, allowing for faster simulations of strongly convective flows. 

The efficiency of LBM over LGA did not come without limitations and the computational cost of highly nonlinear flows needed a huge number of lattice sites, which were out of reach of current high-performance computers. As a result of this bottleneck in the number of lattice sites, a new solution was proposed: quantum computing. Leveraging entanglement and superposition to realize every collision and propagation step for each lattice site all at once with a single process, taking advantage of the locality of the method. Following this idea, several quantum algorithms for LBM surged \cite{budinski2021quantum,doi:10.1142/S0219749921500398,kocherla2024fully,PhysRevApplied.21.034010}. These implementations successfully model Navier-Stokes equations with an exponential speedup. However, due to the linear nature of quantum computing, nonlinear terms in the collision operator restricted the algorithm to a single time step. Measurement to transfer quantum states to classical information and state preparation to load the classical data into the quantum system were needed at each time step. This intermediate step scales linearly with the number of lattice sites, erasing the practical advantage of these algorithms. Additionally, even for the linear terms in LBM, the computation of more than one time step decreases the final probability to measure the correct information into the system, which results in a very inaccurate computation \cite{wawrzyniak2025dynamic}. Given these limitations, four different approaches were proposed to achieve a real quantum advantage in CFD.

The first approach combines classical and quantum computers (hybrid method) to compute linear terms with the quantum computer and nonlinear terms with the classical computer \cite{shinde2024utilizing,shukla2023hybrid}. Under this framework, quantum computing would act as an additional processor (quantum processing unit (QPU)), which only handles specific problems. Nonetheless, for a high number of lattice sites, the classical computing unit (CPU) would act as a bottleneck, while for a low number of lattice sites, the QPU would only slow down the process. This is because quantum computing only provides an advantage for large problems that require a large number of qubits.

Dynamic quantum circuits offer the advantage of mid-circuit measurement, enabling classical logic operations conditioned on the results of those measurements. The second approach consists of using them as a nonlinear and nonunitary operator, which can result into the execution of several time-steps without need for reinitialization.  The applicability of this method is in question due to the complex process of controlling the effects of partial measurements in the quantum state. So far, it has only been successful for linear collision operators \cite{wawrzyniak2025quantum,wawrzyniak2025dynamic}. 

The third approach consists of expanding the nonlinear LBM collision to linear terms by using a Carleman linearization \cite{Carleman1932} explored by Succi and Sanavio \cite{sanavio2024lattice,sanavio2025carleman,sanavio2024three}. In summary, the idea is to consider nonlinear terms as dynamical variables embedded into a large array $r=[x,x^2,x^3,...]$, whose terms will be updated by using other nonlinear terms $x=f(x,x^2,x^3)$, $x^2=f(x^2,x^3)$, etc. Employing a large number of qubits where every quantum state has access to every distribution function $f_i(x)$ and its nonlinear terms, the collision operator can be computed without measurements at each time step. However, the large number of unknowns and the quantum circuit depth that scales as $O(NQ)^4$ with $Q$ the number of channels and $N$ the number of lattice sites, makes the method unfeasible for current quantum computers, even at a second order approximation.

The fourth and last approach consists of using LGA instead of LBM. While LBM has been more successful in the classical realm, quantum computers present a suitable candidate for LGA due to its simplicity and locality. In contrast to LBM, LGA generates the nonlinear terms in the collision operator using deterministic and boolean collision rules \cite{bastida2025efficient,fonio2023quantum} and in some cases has been observed to be more suitable for noise reduction in quantum computing than LBM. Notice that the most well known LGA model for Navier-Stokes equation known as Frisch-Hasslacher-Pomeau (FHP) \cite{PhysRevLett.56.1505} is not deterministic. Nevertheless, the model can be adapted to be deterministic, requiring more time steps and alternating between different collision rules. Other deterministic LGA implementation have also been proposed where Navier-Stokes equations can be modeled \cite{nasilowski1991cellular}.
  
Despite this, an exact algorithm for several time steps has been proven impossible due to the combination collision and streaming operators being nonunitary \cite{Schalkers2024}. Additionally, the high statistical noise of LGA is still present in quantum algorithms, resulting in a worse performance. Another alternative combining LBM probabilities and LGA binary encoding has been proposed, where several time-steps can be computed by increasing the number of qubits linearly with the number of time steps, which translates into an exponential scaling of states \cite{wang2025quantum}.

The Lattice Boltzmann method (LBM) offers a promising avenue for computational fluid dynamics (CFD) due to its inherent locality and suitability for parallelization. Quantum algorithms for LBM hold the potential to overcome its computational bottlenecks. However, realizing efficient quantum LBM implementations, particularly for handling nonlinearities and multi-timestep simulations, remains a significant challenge. In this work, we introduce a novel quantum algorithm that merges the simplicity of LGA  collisions with the equilibrium properties of LBM. This hybrid approach based on Integer Lattice Gas Automata (ILGA) developed by Blommel \textit{et al.} \cite{PhysRevE.97.023310} yields a classical algorithm, operating at the mesoscopic scale, whose computational cost scales with the desired Reynolds number through adjustable collision terms. This approach establishes a connection between LGA/LBM and Molecular Dynamics (MD), allowing us to estimate the probability distribution of molecules and modify their interactions through transition probabilities within the model. Crucially, it eliminates the need for macroscopic variable calculations, relying instead on the distribution of quantum state amplitudes, enabling efficient implementation with $R_y$ gates. Additionally, time-step concatenation by adding qubits or including non-collided terms is possible and nonlinear terms can be handled in the quantum algorithm using tensor products.

The article is divided into two main parts: the classical and quantum algorithms. First, in section II, we introduce the integer LGA model and the necessary changes to obtain a mesoscopic algorithm, and further simplifications. Section III introduces the classical results, consisting of 1D simulations (obtaining equilibrium functions and shockwave equation) and 2D (equilibrium functions, Taylor-Green vortex and lid-driven cavity). In this section, we also discuss how to obtain viscosity, and we compare the computational complexity of LBM and FLGA. Finally, in section IV, we introduce the quantum version of the algorithm and, in section V, the corresponding 1D simulation results. In section VI, we summarize this work and draw conclusions.


\section{From integers to floats in lattice gas automata}
Following the algorithm proposed by Blommel \textit{et al.} on Integer Lattice Gas Automata (ILGA) \cite{PhysRevE.97.023310}, we propose a new variant whose main contribution is to reduce the noise of the model by computing the algorithm 
at a mesoscopic scale. By doing this, additional simplifications are considered to obtain an algorithm with efficiency comparable to LBM. Notice that two algorithms have been proposed for ILGA: a Monte Carlo approach with particle interaction by Blommel \textit{et al.} and a probability sampling method \cite{PhysRevE.105.035303}. In the present article, we work on the collision-based approach, which can be easily simulated at a mesoscopic scale. In this section, we will first review the algorithm proposed by the original authors, and then discuss the contributions of the present article. As the encoding and collision operators are the sole distinctions between ILGA and the new model, we will focus on these aspects. As LBM and LGA basic theory will not be reviewed in this article, we refer the reader to \cite{wolfgladrow2005lattice}.

\begin{figure}[H]
    \centering
    \includegraphics[width=0.5\textwidth]{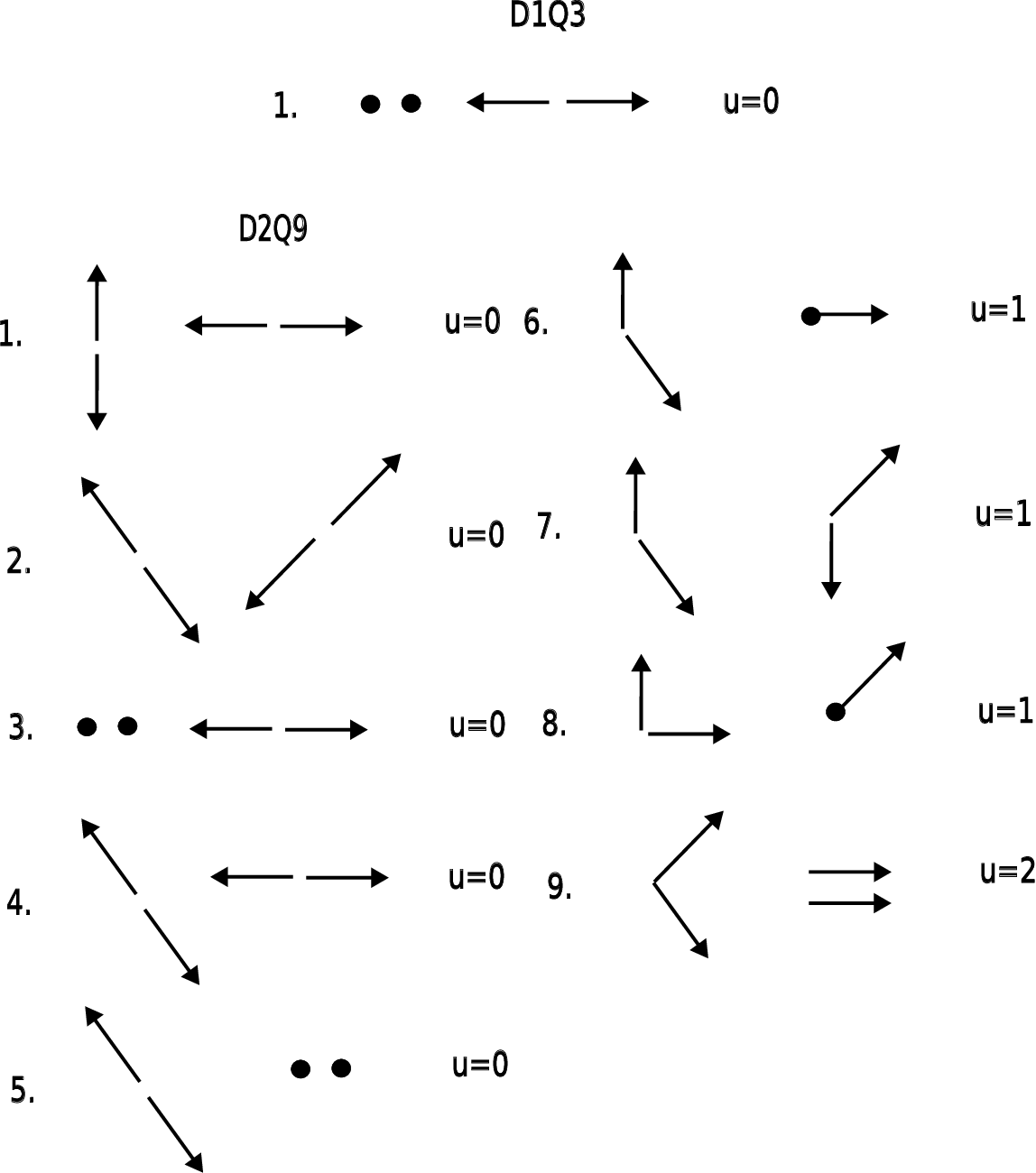}

    \caption{Equivalence classes for D1Q3 and D2Q9 using two-body collisions. Each equivalence class constitutes multiple collisions by 90° degrees rotations or symmetrical mirroring of the collisions displayed.} 
    \label{fig:eqv_classes}
\end{figure}

\subsection{Integer Lattice Gas Automata}
The main idea of the original authors is to produce nonlinear terms by using collisions given by the equivalence classes of LGA, where all particles have the same mass. We define an equivalence class in the context of LGA as the set of collisions that share the same momentum and are related by rotational symmetry using multiples of 90°.  The equivalence classes are defined by the particles whose velocity and mass are the same before and after the collision. For example, in the case of a 1D model with three channels (D1Q3), there is only one equivalence class composed of two rest particles, one particle with right momentum and another with left momentum (each particle having the same mass). The equivalence classes for D1Q3 and D2Q9 with two-body collisions can be seen in Fig~\ref{fig:eqv_classes}. To do so, the authors establish detailed balance, meaning that the forward and backward probabilities are the same. If we define the probability that two particles collide with their channels given by index $i$ and $j$ (each channel with a specific mass and momentum) to obtain other channels with index $k$ and $l$ as $P_{ij \rightarrow kl}$, then by detailed balance

\begin{equation}
p_i p_j P_{ij \rightarrow k,l}=p_k p_l P_{kl \rightarrow i,j}
\end{equation}
where $p_i$, $p_j$, $p_k$, $p_l$ are the probabilities of having one particle in the channels $i$, $j$, $k$, and $l$, respectively. However, to obtain the same proportion of particles at each channel as in LBM, we enforce these probabilities to be equal to the weights of LBM. This condition will lead to a similar equilibrium distribution function as LBM. For D1Q3, the weights are $w=[2/3, 1/6, 1/6]$ for rest, right and left particles. This way, the forward and backward transition probabilities are adjusted with the weights. Additionally, the total probability must be one 

\begin{equation}
    \sum_{k,l} P_{ij\rightarrow kl}=1.
\label{eq:prob1}
\end{equation}
Using these two properties, one obtains

\begin{equation}
    P_{ij\rightarrow kl} =
    \begin{cases}
      \lambda_{ij,kl} \min \left(1, \displaystyle\frac{w_k w_l}{w_i w_j} \right) \delta_{v_ij ,v_kl}& ij\neq kl\\
      1-\sum\limits_{kl\rightarrow ij} P_{ij\rightarrow kl} & ij= kl\\
    \end{cases} 
\label{eq:probtrans}
\end{equation}

where the parameter $\lambda_{ij,kl}\in \mathbb{R}$ is introduced to account for the effective rate at which each collision takes place and guarantee that \eqref{eq:prob1} is fulfilled. It is important to notice that these probabilities are computed beforehand based on the number of channels and targeted viscosity. The relation between viscosity and the effective collision $\lambda$ will be discussed below.

The Monte Carlo approach begins by selecting two random channels to collide. This is done by generating two random numbers $r_1,r_2\in[0,N]$ ($N$ represents the number of particles in a given lattice site) and the cumulative number of particles of the different channels up to the value $r_i$

\begin{equation}
    \sum_{i=0}^{s_1-1} n_i < r_1  \leq \sum_{i=0} ^{s_1} n_i
\end{equation}
with $n_i$ the number of particles in the channel $i$, obtaining the channel $s_1$ to collide for the first random number. A similar procedure is done for the second number. 

Next, we calculate the collision's outcome using the transition probabilities obtained before and a third random number $r\in[0,1]$

\begin{equation}
    \sum_{i=0}^{k}  P_{s_1 s_2\rightarrow (s_3 s_4)_i}<r \leq  \sum_{i=0}^{k+1}  P_{s_1 s_2\rightarrow (s_3 s_4)_i}
\end{equation}

obtaining the pair $s_3 s_4$. The pair order in the summation is arbitrary, containing only the pairs that conserve momentum and mass. 

This process is repeated $C\in \mathbb{Z}$ times, depending on the number of collisions desired in the system, which will have an effect on the viscosity and the mean free path. 

As the authors also mention, the collision operator of LBM $\Omega_i$ is the ensemble average of the one from ILGA $\Xi$. Therefore, the authors calculate the average collision term as 

\begin{equation}
    f_i^{c+1}=f_i^{c}+ \Omega_i= f_i^{c}+ <\Xi_i>
\end{equation}

 where $f_i$ are the local distribution functions obtained from the ensemble average of the number of particles in each lattice site as $f_i=<n_i>$. If we replace this operator by the probabilities given in the previous algorithm, we get

\begin{equation}
    f_i^{c+1}=f_i^{c}+ \sum_{ijkl} \vartheta_i(jklm)\frac{f_j^c f_k^c}{\rho^2} P_{jk\rightarrow lm}
    \label{eq:collision_f}
\end{equation}

with

\begin{equation}
\vartheta_i(jklm)=(\delta_{i,l}+\delta_{i,m}-\delta_{i,j}-\delta_{i,k})
\label{eq:sign_op}
\end{equation}

The equilibrium distribution functions $f_i^{eq}$ are given by $\Omega_i=0$. Solving this equation, the authors obtain

\begin{equation}
    f_i^{eq}(u,\rho)=\rho w_{v_{ix}}\left(1+3v_{ix}u_x+(3v_{ix}-1)(\sqrt{1+3u_x^2}-1)\right)
    \label{eq:f_eq}
\end{equation}

with the 2D distribution functions being given by the tensor product of 1D equilibrium functions 

\begin{equation}
    f_i^{eq}(u_x,u_y,\rho)=f_{x}^{eq}(u_x,\rho)f_{y}^{eq}(u_y,\rho)
    \label{eq:f_eq_2D}
\end{equation}

with $f_{x}^{eq}=f_{0}^{eq}$ if $u_x=0$, $f_{x}^{eq}=f_{1}^{eq}$ for $u_x=1$ and $f_{2}^{eq}$ for $u_x=-1$ and the equivalent for the $y$-direction.

To estimate how fast the moments non-conserved by the collision operator relax, we need to estimate the relaxation time $\tau$. This parameter is related to the rate at which each of the moments relax to their values at equilibrium.  More details about the relaxation time and its relation with LBM can be found in \cite{wolfgladrow2005lattice}.  

Given an arbitrary non-conserved moment $\pi$, we define the deviation from equilibrium after $c$ collisions as $\tilde{\pi^{(c)}}=\pi^{(c)}-\pi^{eq}$, obtaining a relation of the form
\begin{equation}
    \tilde{\pi}^{(c+1)}-\tilde{\pi}^{(c)}=\lambda g(\tilde{\pi}^{(c)},{\tilde{\pi}^{2(c)}})
\end{equation}

with a quadratic dependency as we are using collisions between two particles. However, the BGK approximation for LBM involves only a linear term which yields a simple relaxation time. As specified by the authors, the key here is to consider that for low velocities $u<<1$ and near the equilibrium $\tilde{\pi}<<\rho$, the previous equation can be approximated as a differential equation, leading to

\begin{equation}
    \tau^\pi=\frac{\tilde{\pi}^{(0)}}{\tilde{\pi}^{(0)}-\tilde{\pi}^{(C)}}\approx\frac{1}{1-\exp \left(\displaystyle \frac{-C\gamma(\lambda)}{\rho}\right)}
    \label{eq:relaxation}
\end{equation}

with $\gamma$ a function of the effective rate $\lambda_i$ (for the case there is more than one equivalence class). The specific expression of this rate will depend on each case, being for a D1Q3 model, $\gamma(\lambda)=\lambda$.  The viscosity can be obtained following the relation of viscosity $\nu$ and relaxation time $\tau$ given by LBM

\begin{equation}
    \nu=\frac{(\tau-\frac{1}{2})}{c_s^2}
    \label{eq:viscosity_tau}
\end{equation}

\begin{figure}[H]
    \centering
    \includegraphics[width=0.15\textwidth]{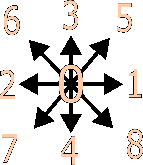}

    \caption{Scheme of channels for D2Q9. Each number represents the index used to map each channel.} 
    \label{fig:d2q9}
\end{figure}

For 2D, the expression of the moments at equilibrium becomes lengthy and cumbersome. To simplify these, the authors approximate $u_x^2\approx0$ and neglect second-order moments in the relaxation time. As shown in \cite{PhysRevE.97.023310}, the authors find the expression

\begin{equation}
    \sigma_{xy}^{(c+1)}=\sigma_{xy}^{(c)}-\frac{1}{9}[8\lambda_1+\lambda_3+2(\lambda_4+\lambda_6+2\lambda_7+4\lambda_8)]\sigma_{xy}
    \label{eq:sigmaxy}
\end{equation}

for the moment corresponding to the shear stress defined as $\sigma_{xy}=3(n_5-n_6+n_7-n_8)$ following the D2Q9 scheme in Fig~\ref{fig:d2q9}. Using the definition of the relaxation time and this approximation, we can get an expression of the relaxation time that is independent of the shear stress

\begin{equation}
    \tau^{\sigma_{xy}}=\frac{1}{C\gamma(\lambda)}
    \label{eq:sigma_tau}
\end{equation}

with 

\begin{equation}
    \gamma(\lambda)=\frac{1}{9}[8\lambda_1+\lambda_3+2(\lambda_4+\lambda_6+2\lambda_7+4\lambda_8)]
    \label{eq:gamma_d2q9}
\end{equation}

\subsection{Float Lattice Gas Automata} 

The model developed by Blommel and Wagner works as intended, recovering the equilibrium functions of LBM and calculating the relaxation time accurately. However, the method is neither efficient nor optimal for general purpose, since it was developed and intended to add noise to LBM. In this section, we will cover some simplifications and changes to make ILGA a model that can efficiently model Navier-Stokes equations as LBM does. 

The first of the simplifications proposed in this article is the change of the model from a stochastic to a deterministic one. The noise source in ILGA comes from the Monte Carlo approach, which consists of a random process depending on the number of particles in each channel and the transition probabilities. 
Instead, we will use floating-point numbers (i.e. floats) to do fractional collisions, directly computing the ensemble average of the integer occupation numbers according to \eqref{eq:probtrans}. This modification will provide the same results but simplifies the computation and eliminates the noise. The collision operator takes simply the form of \eqref{eq:collision_f}. As the collision is equivalent, the relaxation and viscosity are equivalent to ILGA. 

The transition from ILGA to FLGA comes naturally when we think that using a float number (when any fractional value for the number of particles can be chosen) is equivalent to taking the limit of $N\rightarrow \infty$. In such a case, the precision at which we can modify the local distribution functions in the same sense as in LBM $f_i$ tends to 0. 
By directly computing the ensemble average of the collision operator from ILGA, we are transitioning from a model based on microscopic rules to a mesoscopic model, equivalent to the transition from LGA to LBM.
The second simplification introduced here deals with the number of collisions needed to recover the viscosity. As seen in \eqref{eq:relaxation}, the relaxation time depends on the number of collisions $C$ and their effective rate $\lambda$. ILGA restricts the values of $\lambda$, as \eqref{eq:probtrans} uses this parameter to guarantee that \eqref{eq:prob1} is fulfilled. In contrast, in FLGA, we can manipulate the equations to impose detailed balance without the sum of the transition probabilities to be 1. By deleting this probabilistic interpretation, we can directly model the ensemble average of the collision operator $<\Xi>$. The collision operator must be of the same order of magnitude as the local distribution functions to ensure that the collision term is as high as possible to obtain a low viscosity. This can be done by imposing $O(\frac{f_i^c}{\Omega_i})=O(1)$. To this end, we perform an order of magnitude analysis of the collision operator obtaining for $Q$ number of channels  
 
\begin{equation}
    O \left( \sum_{ijkl} \nu_i(jklm)\frac{f_j f_k}{\rho^2} P_{jk\rightarrow lm} \right)=O \left( \frac{\rho^2}{Q^2\rho^2} \right)=O(1)
    \label{eq:order_magnitude}
\end{equation}

The collision operator's order of magnitude is $O(1)$ instead of $O(f_i)$. To change this, we can manipulate the collision operator to include only the local density $\rho$ in the denominator as a linear term, which changes the order of the collision as desired. Therefore, the relaxation of the non-conserved momentum change to

\begin{equation}
    \tau=\frac{1}{1- \exp(-C\gamma(\lambda))}
\end{equation}

which will decrease $\tau$. In the case of interest of LBM at low Mach number and especially for incompressible flows, it is not the density but its increment $\Delta \rho$, which are relevant and therefore, in the model, we interpret $\rho$ as a normalized parameter. Alternatively, the density can be removed from the equation for incompressible flows and substituted as $\rho=\frac{1}{N}$.

A second way to decrease the number of collisions is to modify the influence of the parameter $C$. We interpret the number of collisions with the effective rate $\gamma(\lambda)$ as the total number of effective collisions. However, in FLGA, $C$ can be interpreted as a factor that multiplies $\lambda$ to increase the collision rate and reduce $\tau$, allowing for $\tau\in[0.5,1]$, therefore $C\in \mathbb{R}$. The parameter $C$ is only restricted by the instabilities of the model that occur when we increase $C$ to high values, making the collision operator much larger than the local distribution functions. Notice that at large values of $C$, the non-conserved momentum fluctuations $\tilde{\pi}^{(c)}$ will be significant. This means that $\rho\sim \tilde{\pi}^{(1)}$ for significant collisions, thereby altering the expression for \eqref{eq:relaxation}. These large deviations from equilibrium, which approach the limit of numerical instabilities, can be understood as an over-relaxation of the distribution functions. This over-relaxation enables the simulation of small viscosity flows, effectively bypassing the restriction that enforces $\tau > 1$ in ILGA. Taking into account that we cannot neglect second-order fluctuations, the relaxation time takes the form of 

\begin{equation}
    \tau=\frac{1}{ \displaystyle \tilde{\pi}^0 \left( \exp(\gamma(\lambda) )-\frac{1}{2\sqrt{6}} \right)+\exp(\gamma(\lambda))}-1
    \label{eq:tau_no_simplified}
\end{equation}

for 1D. In the case of a compressible regime where the density may play an important role, we can include a dependency of $C(\rho)$. Therefore, the parameter used to scale each collision will depend on the density, but we will not consider this case as it is out of the scope of this article. 

The final optimization, from ILGA to FLGA, reduces collision costs by simplifying the number of terms in the collision. In the ILGA model, the order of colliding particles within an equivalence class matters. For example, a collision given by $f_i f_j$ differs from $f_jf_i$. The transition probabilities ($P_{i,j \rightarrow k,l}$) for all possible combinations of incoming ($i, j$) and outgoing ($k, l$) particle pairs that conserve mass and momentum must be considered. The sum of these probabilities must also equal one ($\sum_{k,l} P_{i,j \rightarrow k,l} = 1$). This necessitates tracking and calculating numerous combinations.

However, in the FLGA model, particle order is irrelevant. A collision between two particles is treated the same regardless of their original order. This indistinguishability significantly reduces the number of unique collisions that need to be considered, simplifying the computation and making the FLGA model more efficient. This is especially important for 2D. Consider that from around 300 collisions in D2Q9, this can be reduced to 60 due to these simplifications, with only 22 pairs of products to be precomputed at each time step.

\section{Classical implementation results}

\subsection{Equilibrium distribution functions}

\begin{figure}
    \includegraphics[width=0.45\textwidth]{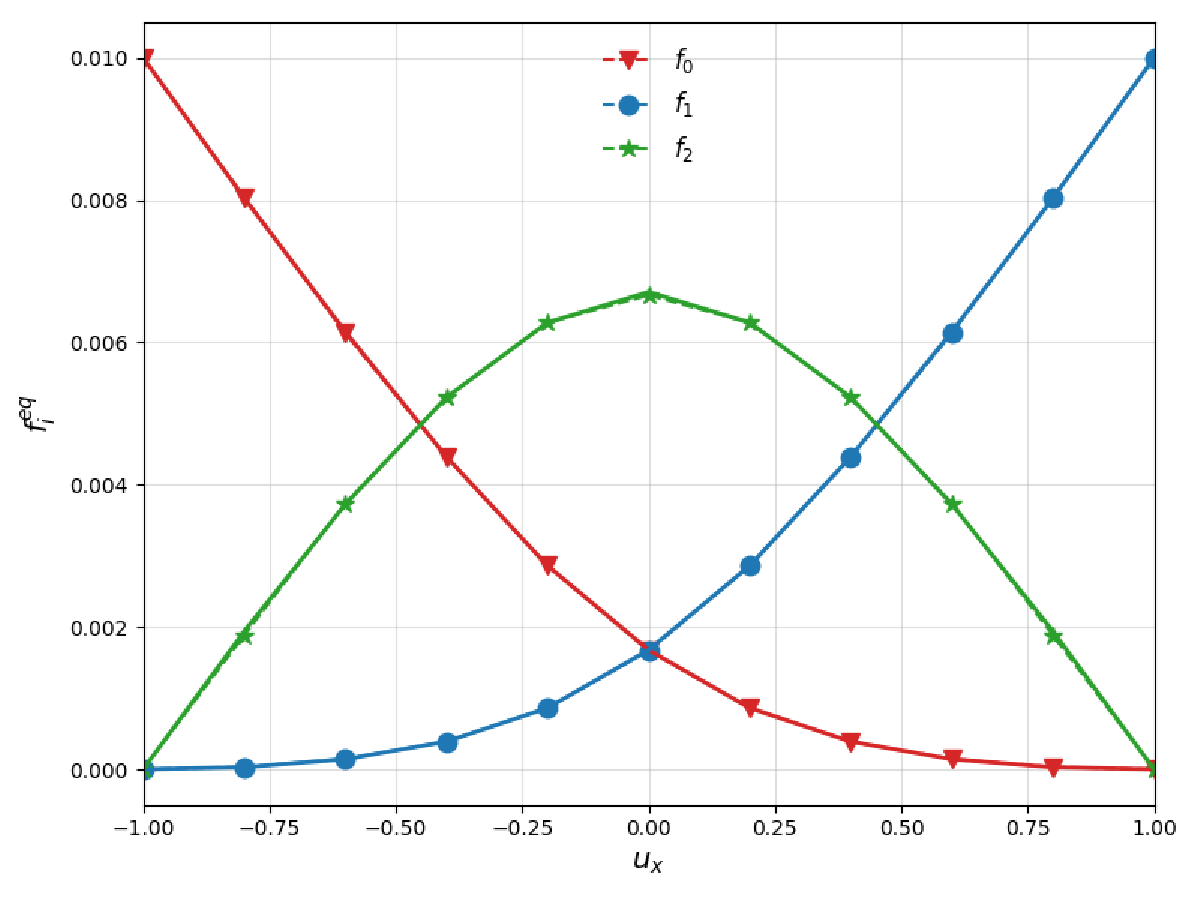}
    \caption{Equilibrium distribution functions of FLGA using a D1Q3 model. The symbol $f_0$ refers to the rest particle, $f_1$ for the right particle and $f_2$ for the left particle. Solid lines refer to the equilibrium distribution functions given by equation \eqref{eq:f_eq}, while the markers are simulation data.} 
    \label{fig:eq_D1}
\end{figure}

Similarly to the ILGA article by Blommel \textit{et al.} \cite{PhysRevE.97.023310}, the first thing to verify is that the equilibrium functions from FLGA are the same as ILGA equations \eqref{eq:f_eq}, which match LBM equilibrium functions for $u=0$. To check this for 1D, we used $\lambda=1.5$ with the transition probabilities $P_{-11,00}=\lambda$ and $P_{00,-1 1}=\frac{\lambda}{16}$. Notice that $\lambda$ can be larger than 1 for FLGA. The number of sites is set to $L=100$ with the number of time steps $T=1000$. The first 500 steps are used to reach equilibrium distributions and only during the last 500 time steps we calculate the macroscopic variables and average the result for the final plot. The initialization of left and right channels has a probability 
\begin{equation}
P(x)=\frac{1\pm U}{2}sin\left( \frac{\pi x} {L}\right)
\end{equation}
with $U=[-1:1]$ in intervals of 0.1 in FLGA units. For each interval, we reinitialize the lattice to obtain different initial velocity. The results can be seen in 
Figure~\ref{fig:eq_D1}, obtaining the same distributions as the equilibrium distribution functions for D1Q3 \eqref{eq:f_eq}, which are the same as LBM equilibrium functions for low velocities. 

\begin{figure}
     \includegraphics[width=0.45\textwidth]{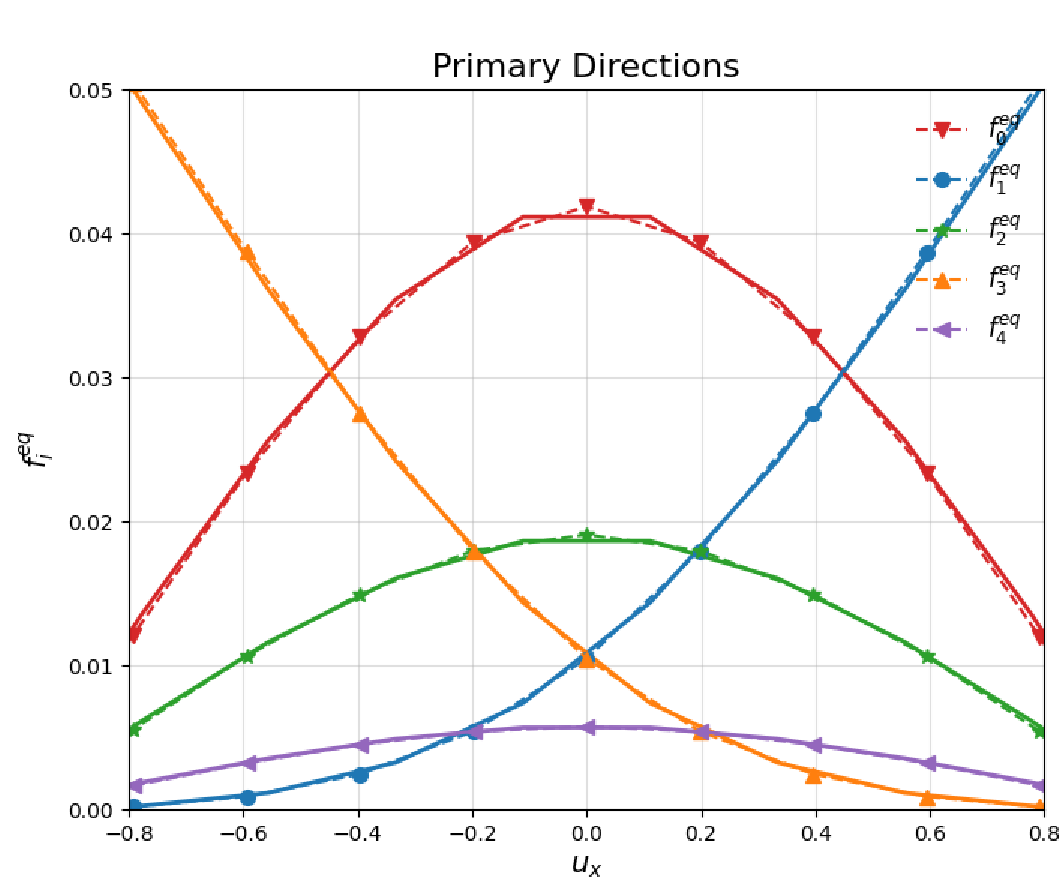}
        \includegraphics[width=0.45\textwidth]{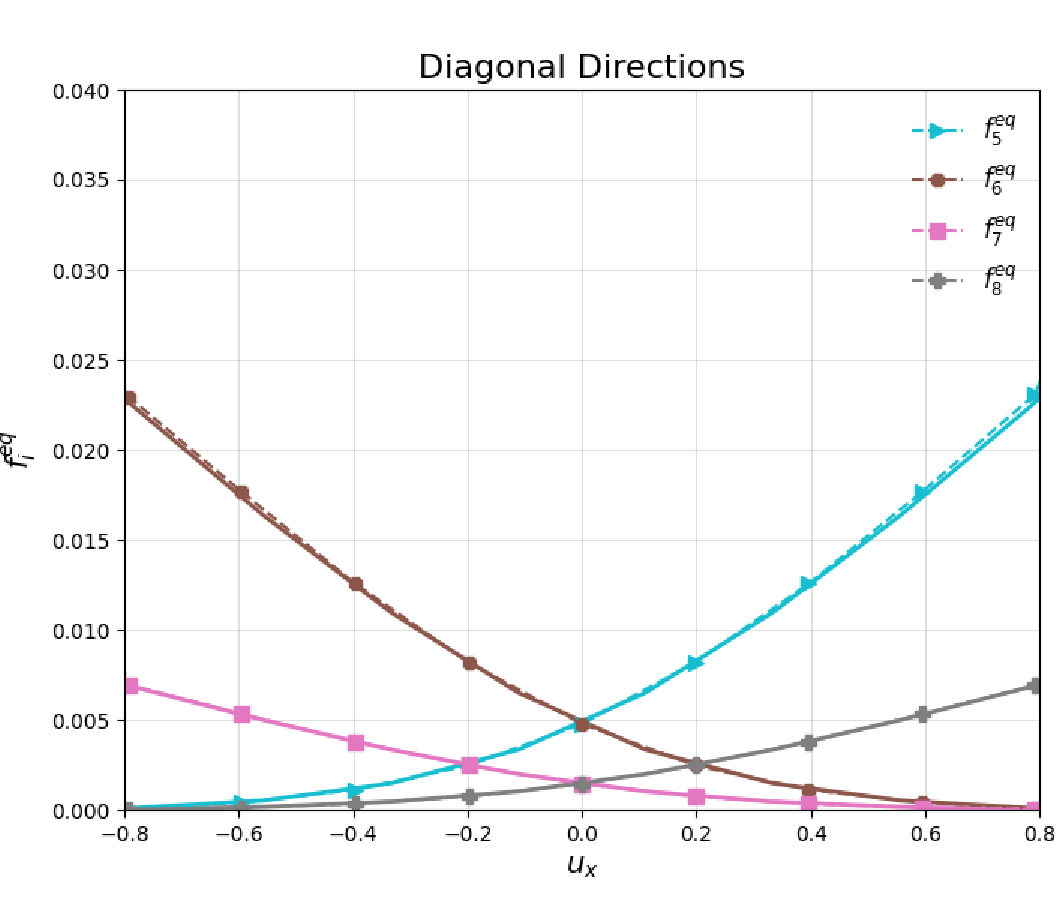}
    \caption{Equilibrium distribution functions of FLGA using a D2Q3 model, with solid lines representing the theoretical equilibrium functions given by \eqref{eq:f_eq_2D} and the dash lines the simulation data. The subindex order follows} 
    \label{fig:eq_D2}
\end{figure}

Something similar can be done in 2D, initializing the lattice with different $u_x$ velocities in each case, obtaining different values of $f_i^{eq}$ for each channel. In this case, we used a lattice with $L_x=15$ and $L_y=15$ sites and $T=400$ steps with 300 steps to reach the equilibrium. We also obtained the same results as the analytical solution suggested, proving that the approximations from ILGA to FLGA maintain accuracy (see Figure~\ref{fig:eq_D2}).

\subsection{D1Q3: Shockwave simulation}

\begin{figure}

    \includegraphics[width=0.48\textwidth]{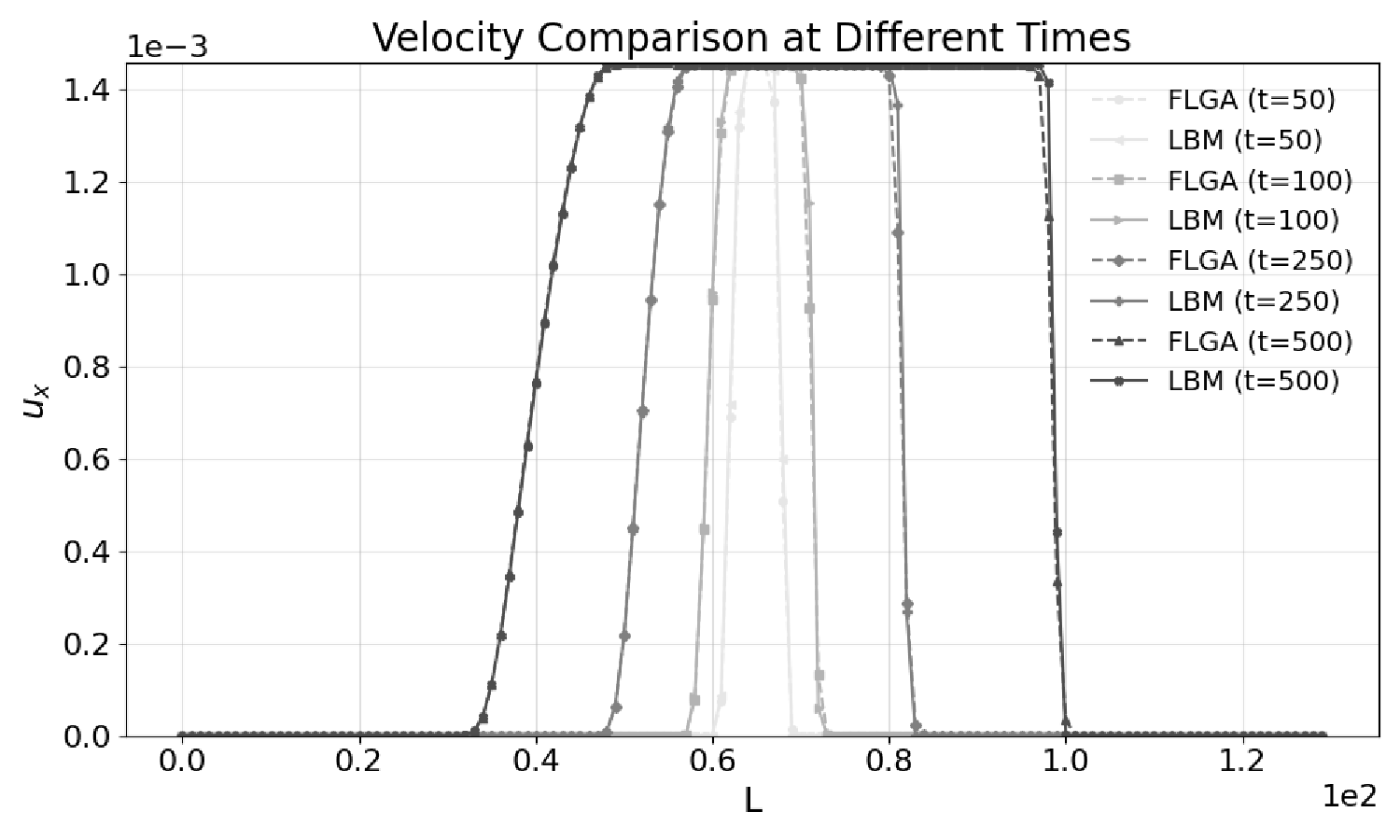}
\includegraphics[width=0.48\textwidth]{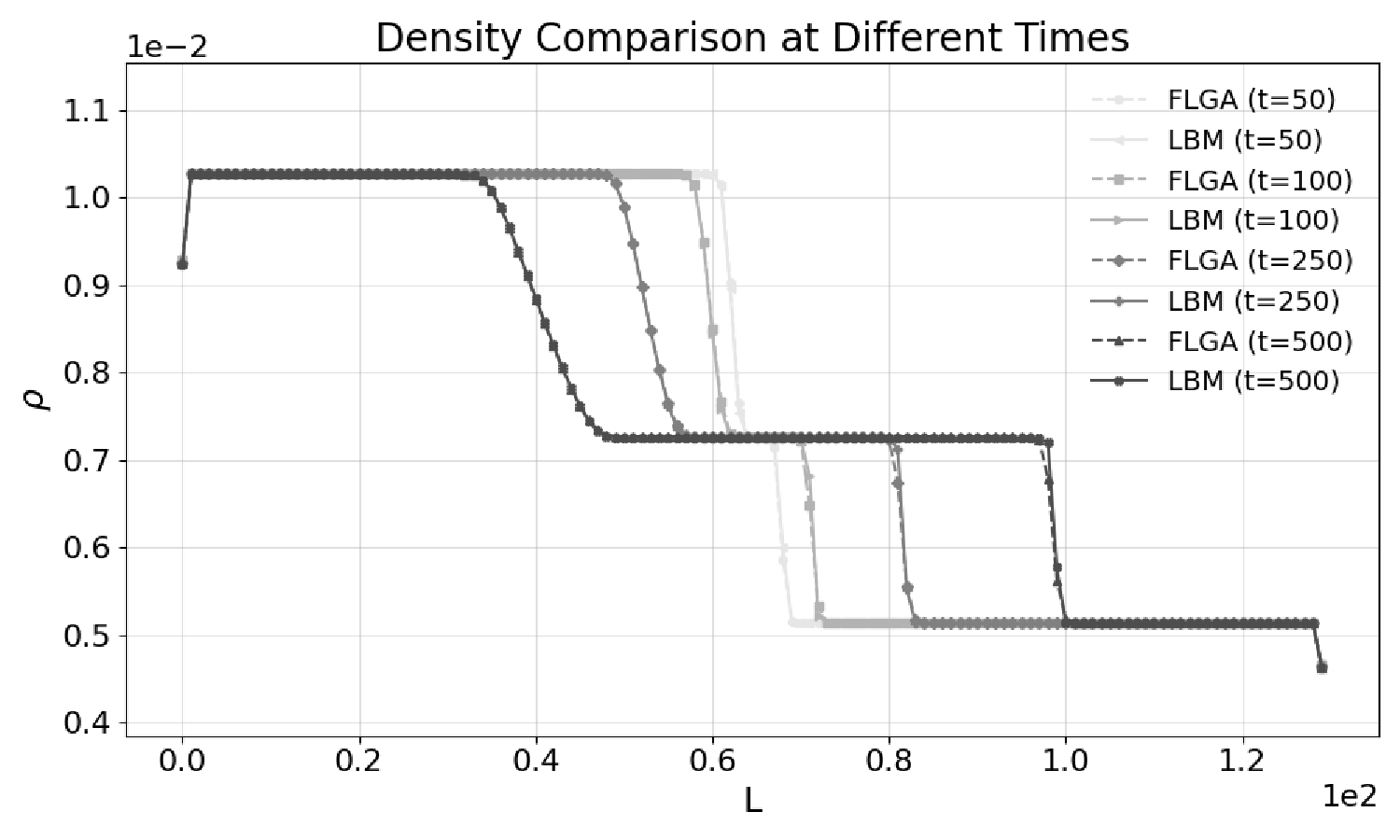}
        
    \caption{Density $\rho$ (upper panel) and velocity $u_x$ (lower panel) for a D1Q3 shockwave model using FLGA with $T=1280$ and averaging per lattice site of 10.  } 
    \label{fig:shockwave}
\end{figure}

Once we have proven that FLGA can recover the equilibrium distribution functions, the next step is to test the model for a nonlinear flow case and compare it with LBM. For that purpose, we have chosen a  1D Riemann problem. The computational setup is the same as in \cite{abdellah2014lattice}: impermeable walls at the two ends of the 1D computational domain with a bounce-back boundary condition. 
The initial velocity is set to zero,  $u_x=0$, while the initial density field is defined as  $\rho_1=4$ and $\rho_2=2$  for the first and second half of the domain, respectively The computed results are displayed in Figure~\ref{fig:shockwave}. In this case, both density and velocity give the same results, with low discrepancies that may come from the difference in the viscosity. While in LBM we have selected the relaxation time $\tau$ via BGK approximation, in FLGA we generate the relaxation time according to the collision rate. However, as $\tau$ is obtained as an approximation that depends on the parameter $\gamma(\lambda)$ (in this case $\gamma(\lambda)=\lambda)$, the real viscosities have some margin of error and may differ. 
Specifically, in this case, we have chosen $\tau_{LBM}=2.1$ and $\tau_{FLGA}\approx\frac{1}{1-exp(-\frac{\lambda}{2})}$ with $\lambda=1.29$, which means $\tau_{FLGA}\approx \tau_{LBM}$. As we know, this is just an approximation, but for the case of 1D, the moment deviation $\tilde{\pi}$ is not large enough to affect the result significantly.

\begin{figure}
    \includegraphics[width=0.48\textwidth]{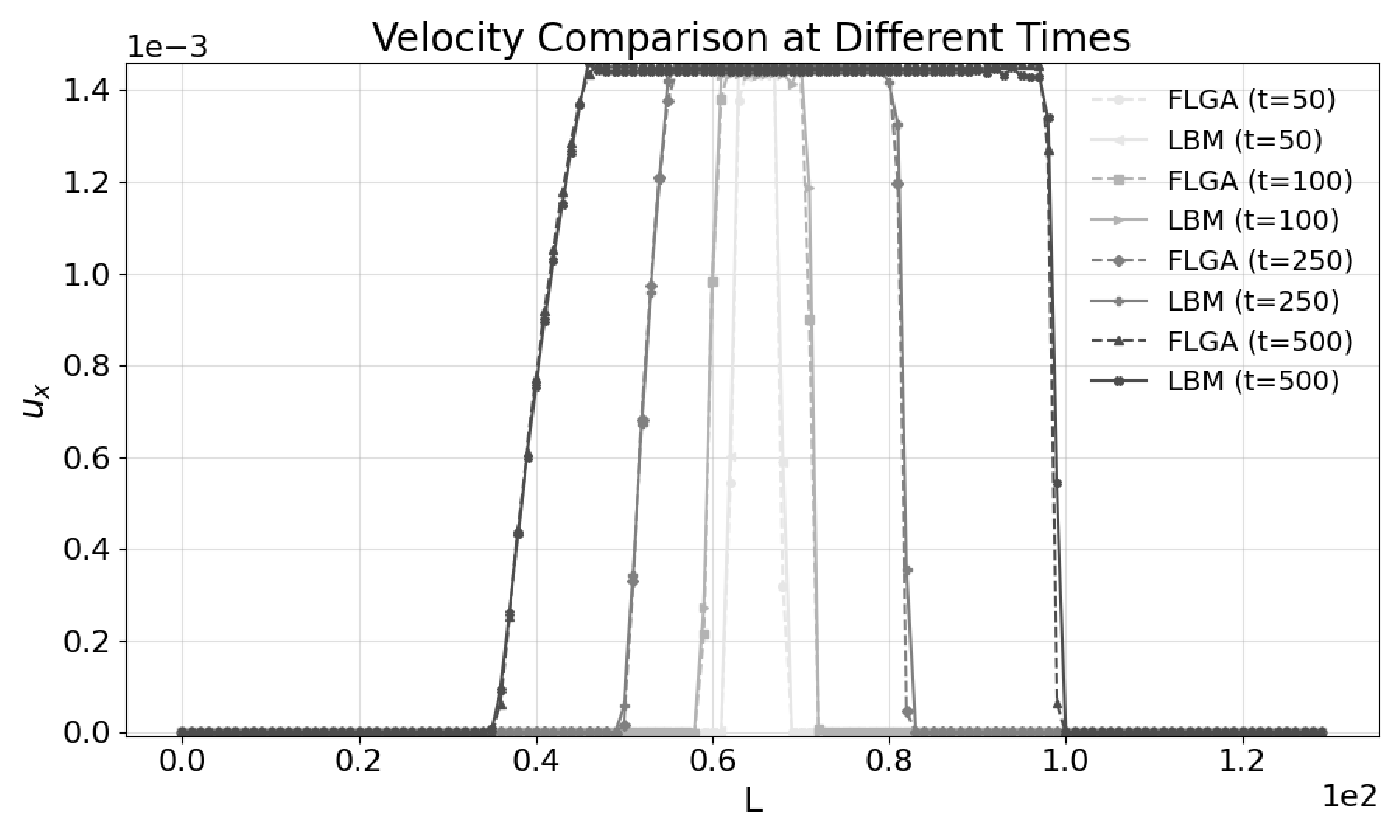}
\includegraphics[width=0.48\textwidth]{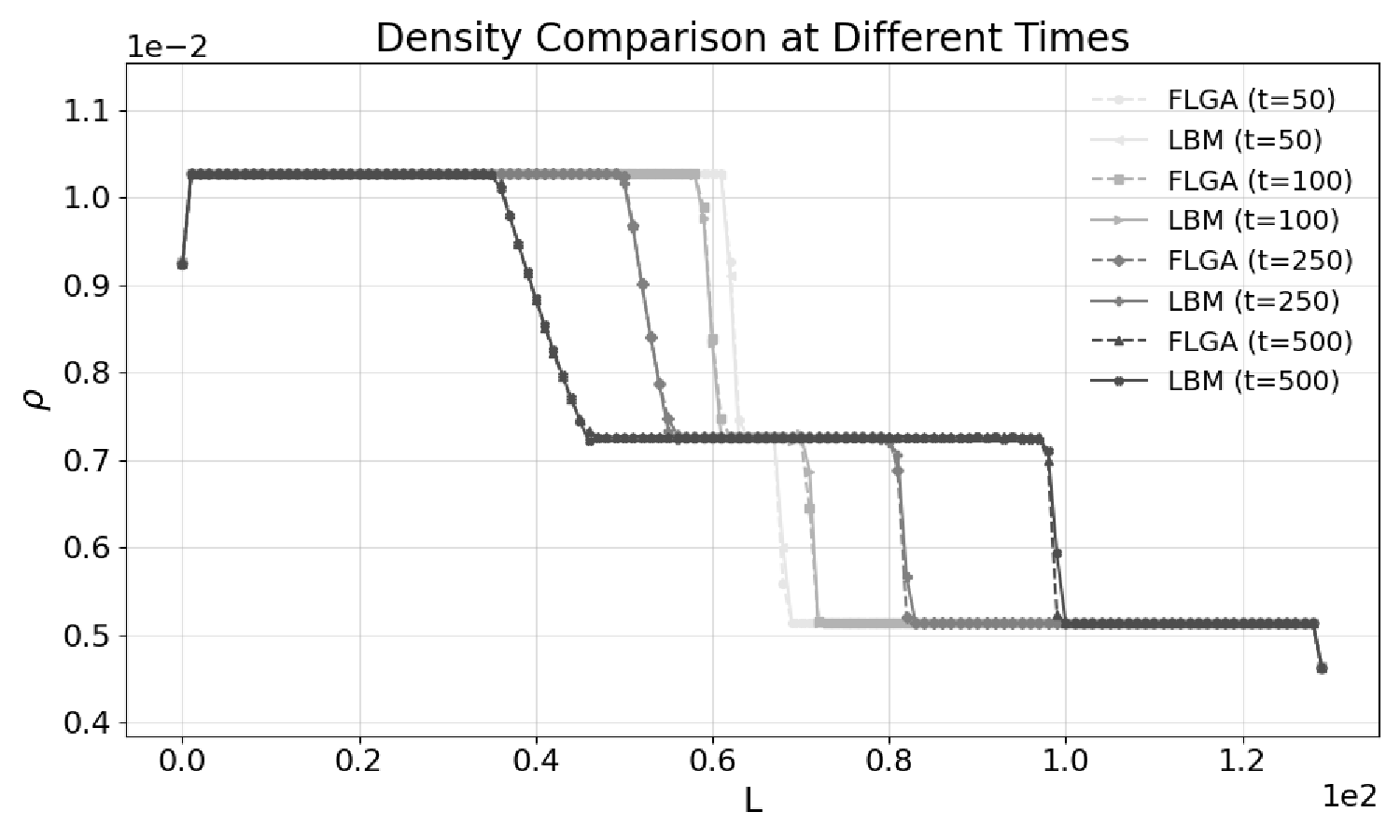}
        
    \caption{Density $\rho$ (upper panel) and velocity $u_x$ (lower panel) for a D1Q3 shockwave model using FLGA with $\tau=0.51$ (over-relaxation), $T=1280$ and averaging per lattice site of 10.} 
    \label{fig:shockwave_overrelaxation}
\end{figure}

Another interesting case, further explored in Sec~\ref{sec:relax}, is the over-relaxation of the distribution functions when $\tau\in[0.5,1]$. In contrast to LBM equiped with a BGK collision operator, where $\tau$ is an assigned value, in ILGA, it is not possible to achieve, as the deviations of the non-conserved moment $\tilde{\pi}$ were much smaller than the density $\rho$. However, in FLGA, this is not the case anymore, and we can use an arbitrary value $C$ that multiplies the effective rate $\lambda$. The theoretical value of $C$ that is equivalent to over-relaxation can be estimated using \eqref{eq:tau_no_simplified}, by calculating the deviation during the simulation or other alternatives, more of this in Sec~\ref{sec:relax}.
 In the Figure~\ref{fig:shockwave_overrelaxation} we can see the comparison of the shockwave for FLGA with LBM but in this case using $\tau=0.51$ and $C=2.7$. This proves that FLGA can also be used in the over-relaxation regime, as we can accurately replicate LBM results.

\subsection{D2Q9: Taylor-Green vortex and Lid-driven cavity}


Given that only theoretical arguments for the applicability of ILGA to more complex cases have been given, we use the FLGA to address two more cases in 2D. Specifically, the Taylor-Green vortex and Lid-driven cavity are standard tests for 2D Navier-Stokes simulations. 

The Taylor-Gren vortex \cite{taylor_green_1937} is a vortical flow described in 2D that has the analytical solution

\begin{align}
    u_x(t) &= -u_{\text{max}} \sqrt{\frac{k_y}{k_x}} \cos(k_x x) \sin(k_y y) e^{-2\nu t}, \\
    u_y(t) &= u_{\text{max}} \sqrt{\frac{k_x}{k_y}} \sin(k_x x) \cos(k_y y)e^{-2\nu t}, \\
    \rho(t) &= \rho_0 + 3P(t), \\
    P(t) &= -\frac{1}{4} u_{\text{max}}^2 \left( \frac{k_y}{k_x} \cos(2k_x x) + \frac{k_x}{k_y} \cos(2k_y y)e^{-4\nu t} \right).
    \label{eq:taylor_green}
\end{align}

\begin{figure} 

    \hfill 
    \begin{subfigure}{0.5\textwidth}
        \includegraphics[width=\linewidth]{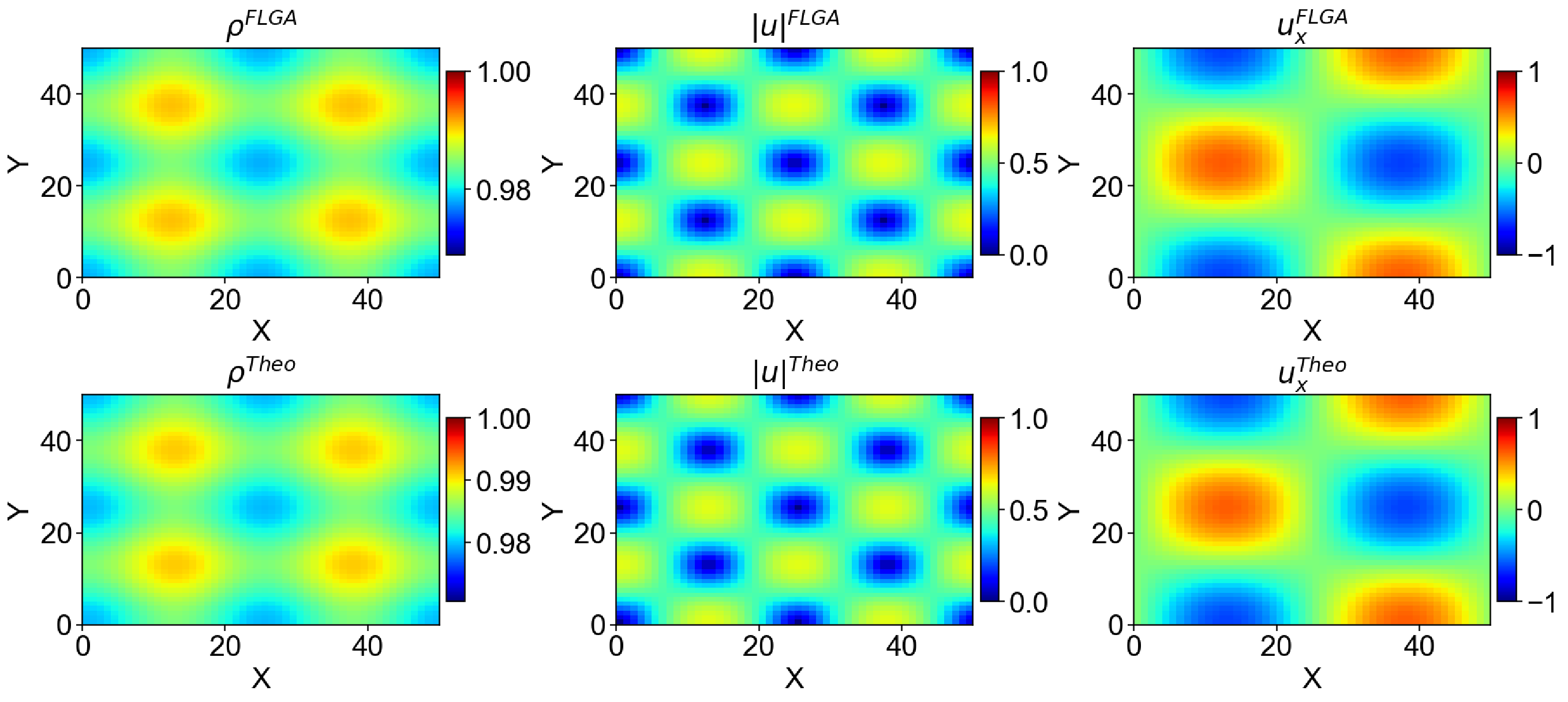} 
        \caption{T=15}
        \label{fig:subfig2}
    \end{subfigure}
    
    \vspace{0.5cm} 

    \begin{subfigure}{0.5\textwidth}
        \includegraphics[width=\linewidth]{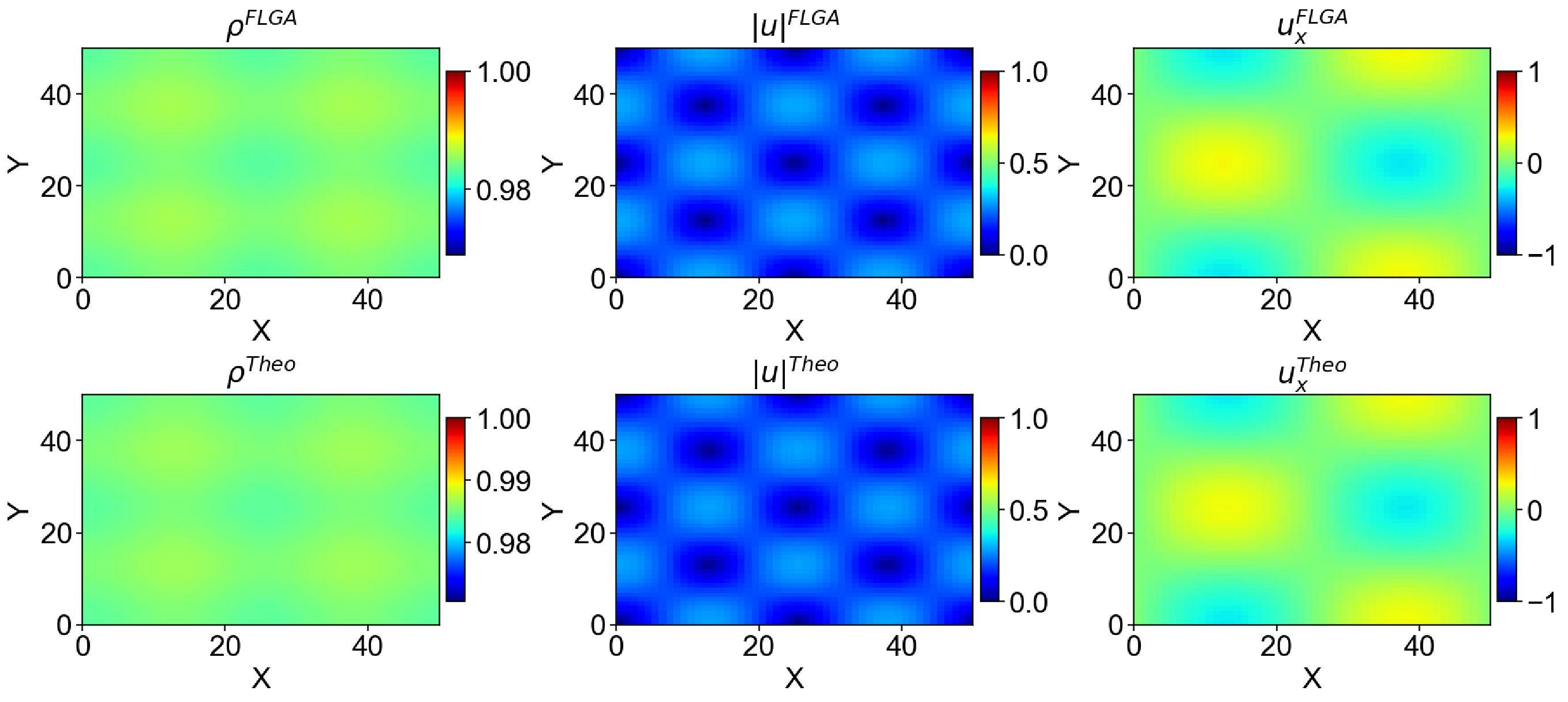} 
        \caption{T=45}
        \label{fig:subfig4}
    \end{subfigure}
    
    \caption{Taylor-Green vortex simulation at different time steps comparing FLGA with the theoretical expression. The figures show the density $\rho$, $X$-velocity $u_x$ and total velocity $|u|=\sqrt{u_x^2+u_y^2}$ from left to right. The simulation relaxation time $\tau=12$ was obtained with $C=0.3$.}
    \label{fig:taylor_green}
\end{figure}

To compute the Taylor-Green vortex decay, we extend FLGA to a D2Q9 model using the equivalence classes shown in Fig~\ref{fig:eqv_classes} (two-body collisions) and setting the initial conditions following \eqref{eq:taylor_green} with $t=0$ (Figure~\ref{fig:taylor_green}).

\begin{figure}
    
    \begin{subfigure}[b]{0.48\textwidth} 
        
        \includegraphics[width=0.49\textwidth,height=4.2cm]{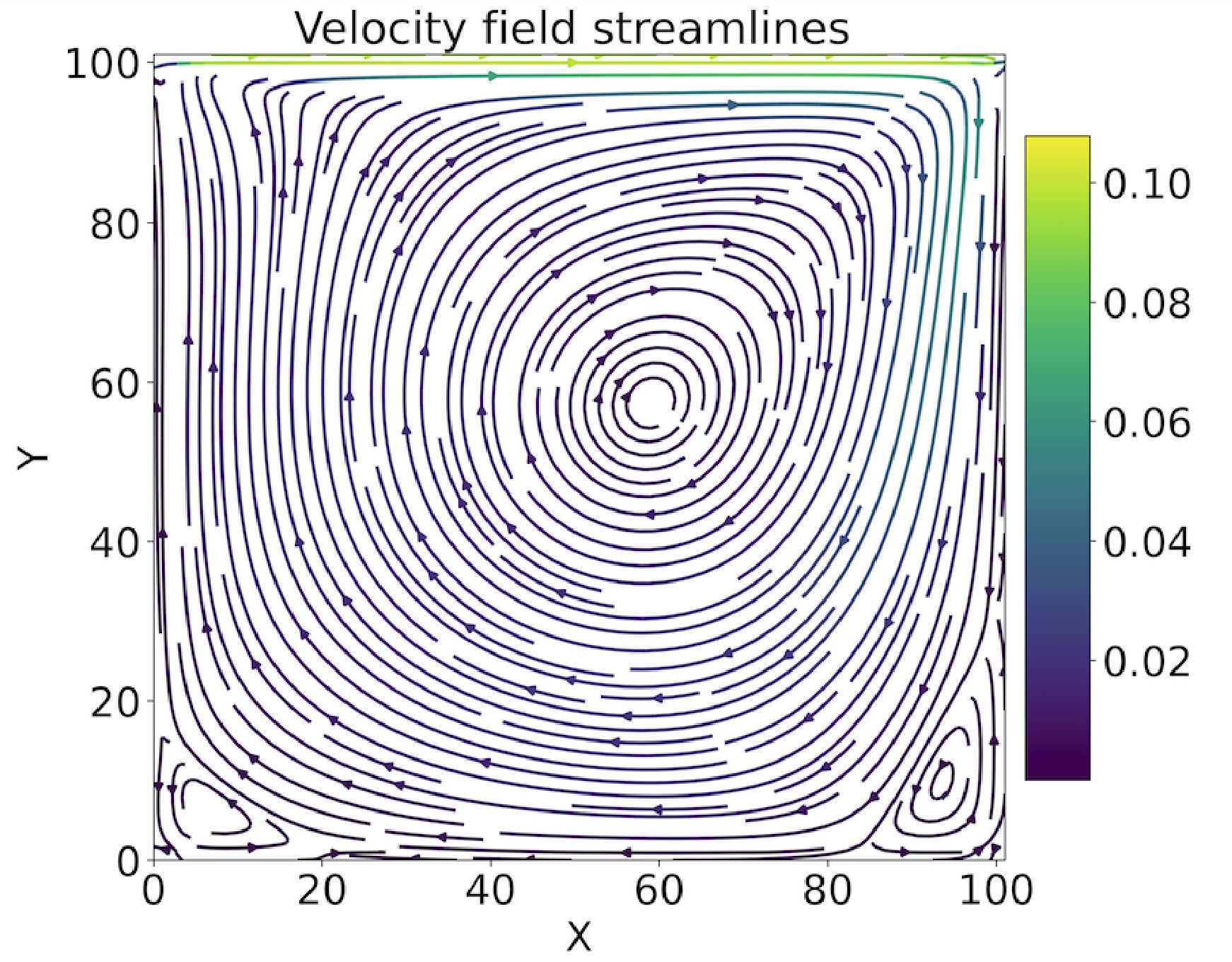}
        \hfill 
        \includegraphics[width=0.49\textwidth]{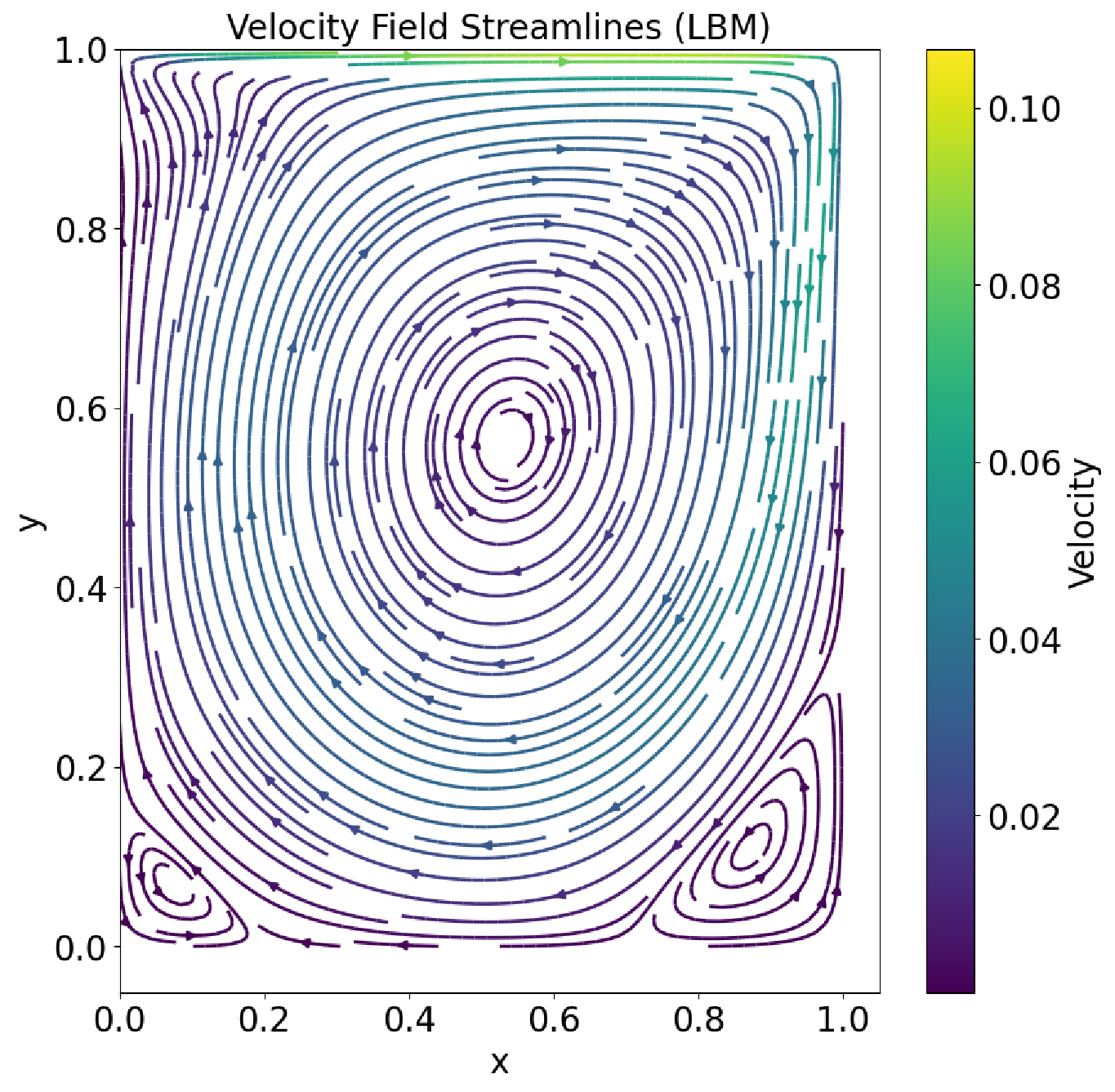}
        \caption{Comparison of the streamline velocity field between FLGA (left) and LBM (right)} 
        \label{fig:subfig1}
    \end{subfigure}
    \hfill 
    \begin{subfigure}[b]{0.48\textwidth} 
        
        \includegraphics[width=0.98\textwidth]{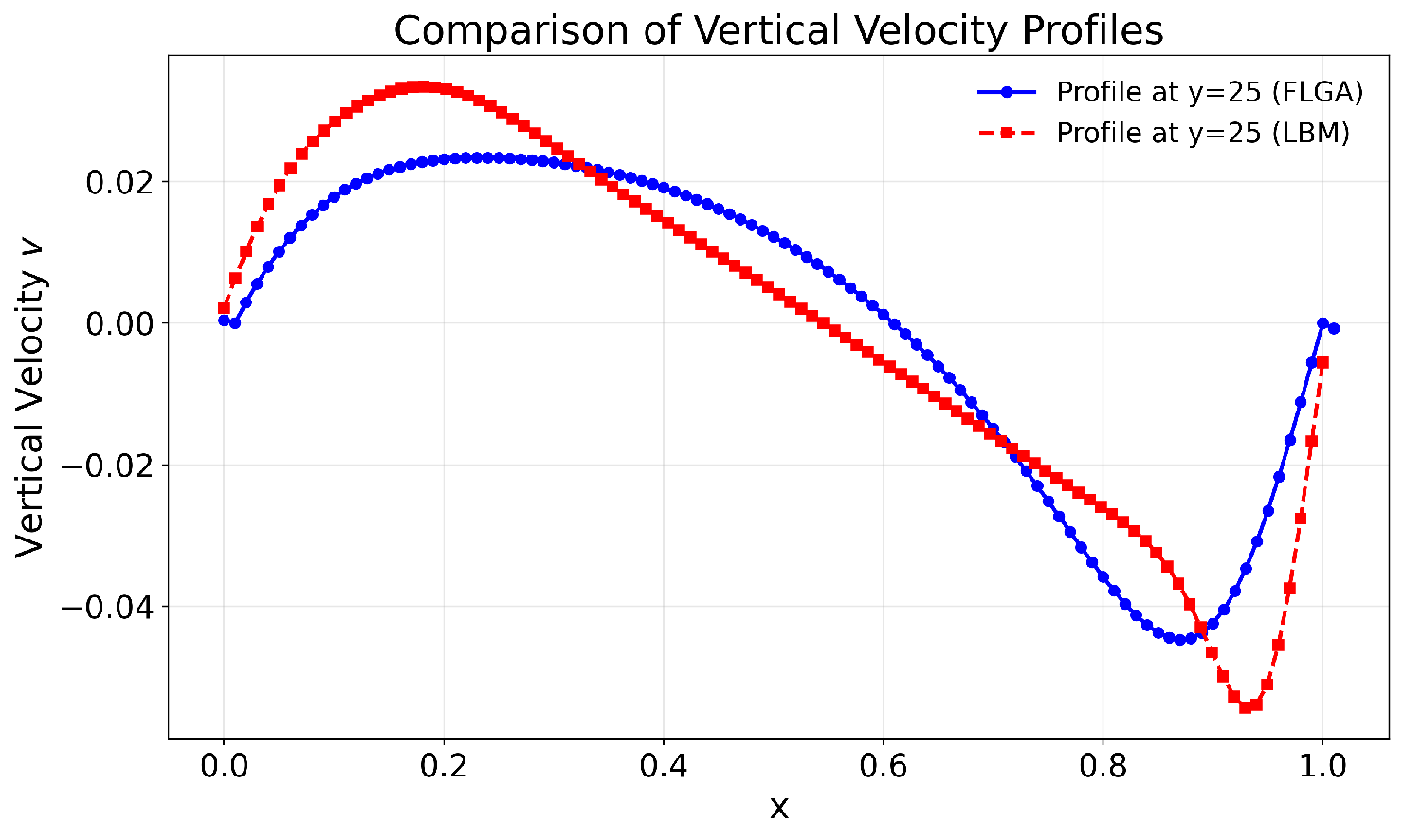}
        \caption{Comparison in the $y$-velocity component between FLGA and LBM at $y=25$.} 
        \label{fig:subfig2}
    \end{subfigure}

    \caption{Lid-driven cavity test using FLGA, where density $\rho$, $x$-velocity $u_x$, and velocity field streamlines are shown compared with LBM. The simulation for FLGA time is $T=20000$ with a factor $C_2=0.2$ and $C_3=1.23$, lattice sites $L=100$, and $u_x=0.2$ for the upper wall. For LBM, similar parameters are chosen, setting the Reynolds number to $Re=1100$.}
    \label{fig:lid_driven}
\end{figure}

The final assessment of the FLGA's reliability for 2D Navier-Stokes simulations was performed using the lid-driven cavity test.  This standard benchmark in CFD consists of a square domain where the top wall moves with a velocity $U$, and the remaining three walls (left, right, and bottom) are stationary.  As shown in Figure~\ref{fig:lid_driven}, the steady-state solution obtained with the FLGA is qualitatively similar 
for the streamlines to that obtained with LBM at a Reynolds number of $Re = 1100$. 
On the other hand, the vertical velocity profile differs due to the difference in the viscosity between both models and their different time to reach the steady state.

However, direct quantitative comparison with LBM or other published results proves challenging due to precisely matching the relaxation time $\tau$ at low viscosities.  The moving upper wall adds momentum into the system, potentially altering the effective viscosity $\nu$ within the domain. Furthermore, for this specific scenario, achieving the low viscosity required to observe the characteristic eddies at the corners was insufficient with only two-body collisions.  Including only three-body collisions did not yield accurate results. A combination of two- and three-body collisions was necessary, further complicating the precise determination of the effective viscosity. This is clear evidence that including boundary conditions affects the relaxation of the different non-conserved moments in the system, requiring further studies. A comparison of $\tau$ and $C$ for two and three-body collisions without boundary conditions can be found in Sec~\ref{sec:three_two}.

A simple estimation based on the relaxation time from Figure~\ref{fig:two_three} yields $\tau = 0.65$, corresponding to a kinematic viscosity of $\nu = 0.05$ and a Reynolds number of $Re = 400$.  However, a visual comparison of the FLGA simulation with LBM results suggests a closer match to $Re = 1100$. Despite the difficulties in estimating the right viscosity when combining collisions with different numbers of channels, we obtain feasible results.

\subsection{Relaxation time: Estimation and dependence with the number of equivalence classes}
\label{sec:relax}
The relaxation time can be difficult to estimate in FLGA. In contrast to LBM under the BGK approximation, where $\tau$ is added as a parameter to obtain the desired viscosity, in FLGA this value is the result of collisions. As in LGA, the higher the number of channels and collisions in the system, the lower the relaxation time. As derived in \cite{PhysRevE.97.023310}, the relaxation time can be obtained as an approximation for $\tilde{\pi}<<\rho$ following \eqref{eq:relaxation}. This approximation works well in ILGA as the collisions are of lower order than the distribution functions. In the case of FLGA, this is not the case anymore, and this approximation cannot be done in general for high values of $\gamma(\lambda)$. The value of the moment deviations will depend on the particular initial and boundary conditions. Notice that in FLGA, the relaxation time is generated in each lattice site and fluctuations are to be expected, as the deviations from the equilibrium at each lattice site and each time step will depend on the local $\pi$. While a better procedure may be possible, in this article, we propose estimating the relaxation time by executing the simulations with a low number of lattice sites, as $\tau$ is not dependent on this value. The relaxation time $\tau$ can be obtained heuristically by comparing LBM with FLGA or using other methods, such as machine learning, when the theoretical expression is unknown. Alternatively, knowing the dependence of $\tau$ and $\pi$, the second-order moment can be computed along the simulation and $C$ adapted dynamically accordingly. This is a common disadvantage of LGA models, where a viscosity calibration depending on the collisions is needed for each case.  

\begin{figure}
    
\includegraphics[width=0.48\textwidth]{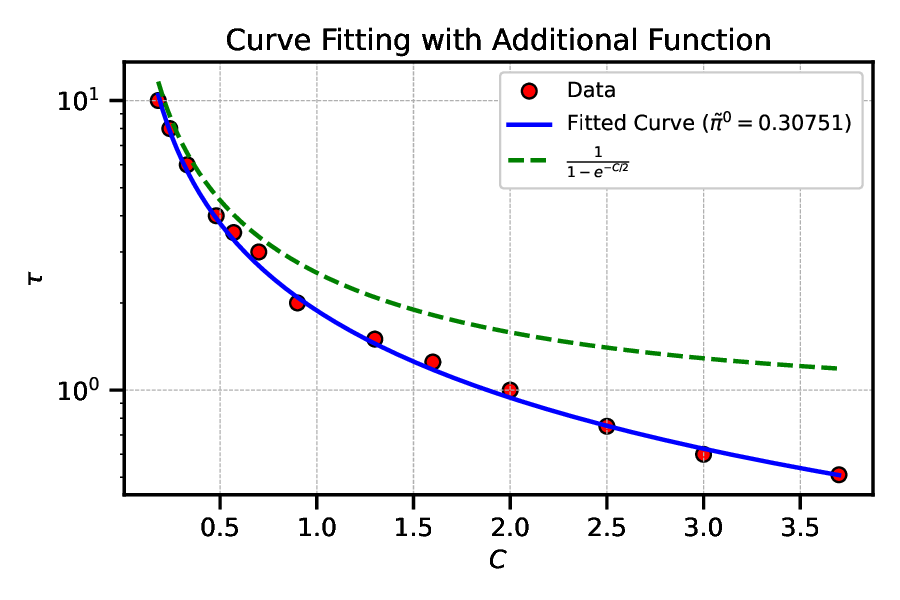}
        
    \caption{Relation between the relaxation time $\tau$ and $C$ given $\lambda=0.5$ for a D1Q3 shockwave simulation. The value of $\tau$ is obtained heuristically comparing LBM with different $\tau$ with the simulation results. The blue line corresponds to a fit of $\tilde{\pi}^0$ in \eqref{eq:tau_no_simplified} and the green line to the result given by the approximation \eqref{eq:viscosity_tau}.}
    \label{fig:1d_tau}
\end{figure}

According to Figure~\ref{fig:1d_tau} for D1Q3, the approximated value fits the simulation data only when $C$ is low and therefore $\tau>1$. For lower relaxation time values, the moment deviations $\tilde{\pi}$ begin to be large and the exact equation needs to be computed. Notice that we have fitted the blue curve in the figure considering a constant value for $\tilde{\pi}$.  This result suggests that given the relation of  $\tau$ with $\pi$, only a few time steps are enough to compute the expected relaxation time, given $C$ (input parameter).

A similar relation can be obtained with a more complex flow such as Taylor-Green vortex in 2D. For this particular case we have compared two different sets of $\lambda$. As suggested in \cite{PhysRevE.97.023310}, each $\lambda$ can be associated with a different equivalence class. For D2Q9 with two-body collisions, there are 9 equivalence classes, as seen in Figure~\ref{fig:eqv_classes}. The authors of \cite{PhysRevE.97.023310} suggested the usage of $\lambda_o=[15/128,1/4,1/4,1/4,1/4,1/4,1/4,1/8,1/8]$. This particular selection aims to avoid the coupling of different moments and prevail the isotropy of the stress tensor. Nonetheless, this particular set can be substituted for any other without observing macroscopic differences, but an improper selection of $\lambda_i$ affects the stability of the system. Following a heuristic approach, we found that the numerical stability improves when all $\lambda_i$ are similar, as $C$ can be increased further without obtaining negative probabilities or other numerical errors. The reason why this happens may be due to the similar collision rate of each equivalence class, avoiding to overcollide some states, obtaining numerical errors such as negative values of $f_i$. Subsequent research should address it in detail. 

\begin{figure}
    
\includegraphics[width=0.48\textwidth]{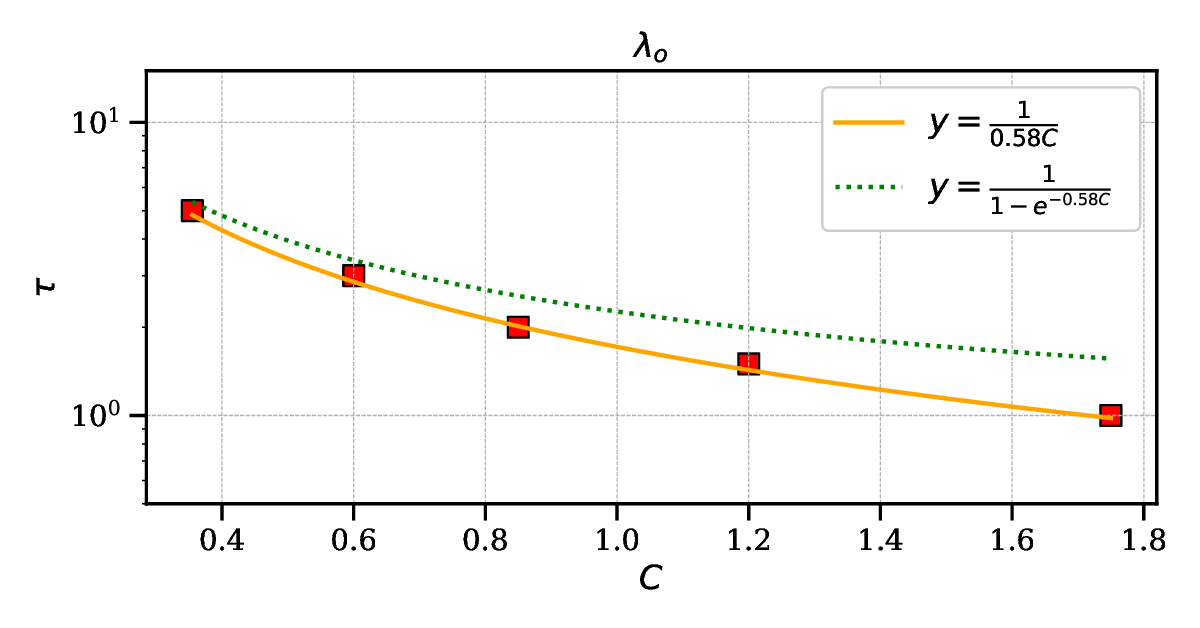}
        \includegraphics[width=0.48\textwidth]{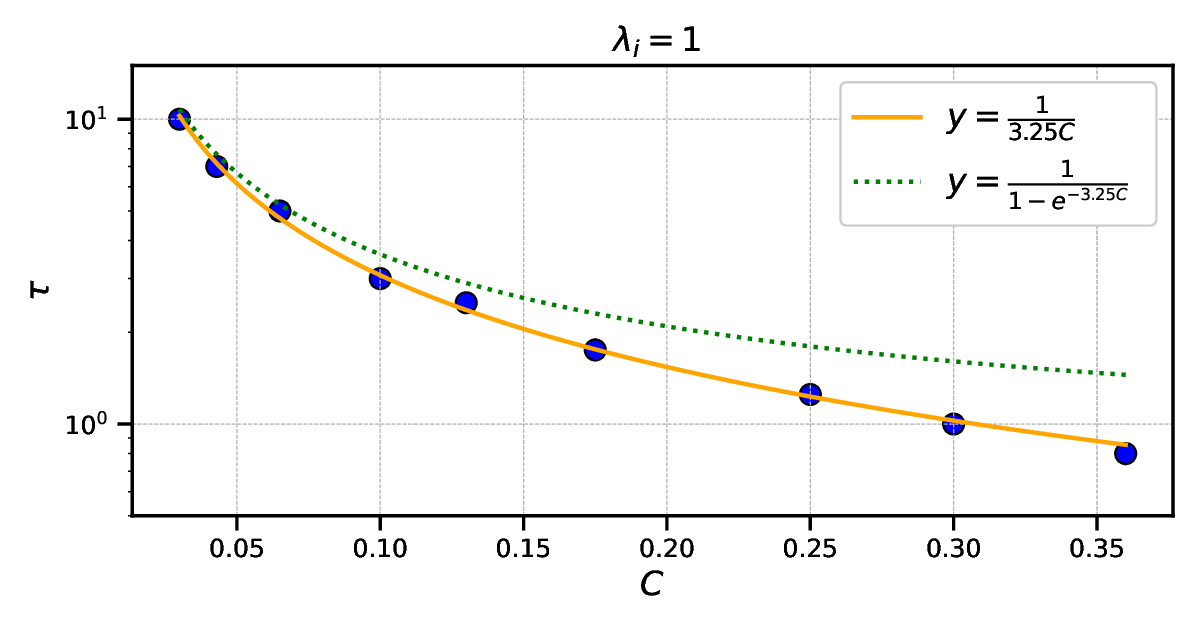}
    \caption{Relation between the relaxation time $\tau$ and $C$ given by two sets of $\lambda$ for a D2Q9 on a Taylor-Green case. $\lambda_o$ refers to the values of $\lambda$ in the original work of ILGA, with $\lambda=[15/128,1/4,1/4,1/4,1/4,1/4,1/4,1/8,1/8]$, while $\lambda_i=1$ refers to all values being 1.}
    \label{fig:2d_tau}
\end{figure}

In Figure~\ref{fig:2d_tau}, we can see how a lower $\tau$ can be achieved when all values of $\lambda$ are similar.  
As we see, the orange line corresponds to a fit of $\tilde{\sigma_{xy}}^0$ considering $u\approx 0$ and neglecting quadratic terms of moments and the definition of relaxation time as $\tau=\frac{\tilde{\sigma}^0}{\tilde{\sigma}^0-\tilde{\sigma}^1}$, while the green line to the result given by the approximation \eqref{eq:viscosity_tau}. The graph shows that when $\lambda$ has similar values, in this case all one, $\tau=0.85$ as its minimum value, while for $\lambda_o$, the minimum value is $\tau=1$. This shows that the stability is better with the first case. Notice that the $C$ values used for each set of $\lambda$ correspond to the interval without numerical errors. For this case, the fitted curve uses $\gamma(\lambda)=0.583$, while the \eqref{eq:sigmaxy} suggests a lower value of $\gamma(\lambda)=0.49$. In contrast, for the case with $\lambda_i=1$, $\gamma(\lambda)=3.25$ has been used while the theoretical value is 2.77. This suggests that the approximation neglecting second order moments and considering low velocities are not accurate for $u>0.05$, decreasiLidng it further with respect to the theoretical approximation. Therefore, we recommend using more advanced tools, such as machine learning, for calibration with LBM. Specifically, a model can learn the relation between tau and C using simulations with few lattice sites, using LBM simulations where we know the viscosity as training data. 
Notice that the relations obtained in Fig~\ref{fig:1d_tau} and Fig~\ref{fig:2d_tau} are valid for any system without boundary conditions, with small differences due to the cross-products of moments in the analytical solution. This is not necessarily the case for cases with boundary conditions where the momentum is not conserved, and the fitted value found may be very different. However, the general inverse relation between $C$ and $\tau$ prevails as the cross products of moments are small and the velocity is near zero. A deeper analysis on how boundary conditions affect $\tau-C$ relation will be studied in future research.

\subsection{Computational complexity and performance}

The computational complexity of FLGA for pair collisions is the same as in LBM of $O(NQ)$ with $Q$ the number of channels and $N$ the number of lattice sites. Even though FLGA requires to compute all two-body collisions or higher orders, which are $O(Q^n)$ with $n$ the maximum number of particles to collide in each equivalence class, these matrices are sparse, not computing most of the terms in \eqref{eq:collision_f}. While for a D1Q3 model, there are only two possible collisions, for D2Q9 with two-body collisions this increases to 60 with 33 different pairs from the 91 possible ones and using three-body collisions, the number of terms in the collision operator increases to 156 with 77 different triplets to compute from the 729 possible ones. 
For four-body collisions, we find only six quadruplets to compute and six possible collisions. To be more precise, the FLGA collision rule is composed of three main steps, all of which scale similarly. 
First, we need to compute the different pairs, triplets, quadruplets, and others that need to be used. Notice that in most cases, during this article we have used two or three-body collisions, specified in each case. Second, the scaling factor depending on $\lambda$, $C$ and $w_i$ is multiplied with the computed product of local distribution functions, and third, these are summed or subtracted according to \eqref{eq:sign_op}. The computational complexity can be approximated as  $O(Nk(\frac{Q}{k}))$ with $k$ for $k$-body collisions where   $(\frac{Q}{k})$ is combinatory of $Q$ over $k$. The higher $k$ is, the larger is the reduction compared with $(\frac{Q}{k})$ due to the restriction of imposing the conservation of momentum. On the other hand, LBM scales as $O(NQ)$.  

In terms of memory, LBM requires saving each local distribution function $f_i$ with $O(NQ)$, each equilibrium distribution function $f_i^{eq}$ with $O(NQ)$, the velocity $u$ with $O(N)$ and density $\rho$ with $O(N)$ (for compressible case). In contrast, for FLGA, we have $f_i$ with $O(NQ)$, the parameters to multiply the distribution functions $O((\frac{Q}{k}))$ and the k-collision $O(N)$. While this depends on the particular implementation of each algorithm, there is an argument to consider that FLGA may be more efficient in terms of memory. LBM requires the benefit of vectorized registers to do operations in parallel. In specific, this is applied to the calculation of velocities, densities, and their products to obtain the equilibrium functions. Otherwise, the computation would be expensive as many read-and-write instructions would emerge. FLGA does not require calculating velocities, and the density calculation can be avoided by setting a larger $C$. This means that for FLGA, the computation requires less memory and has easier data access since the computations are directly in local distribution functions. In LBM, a similar procedure can be implemented, but all the product of distribution functions contribute to the velocities and the final $f_i^{eq}$, which requires more data.

\begin{figure}
    
\includegraphics[width=0.48\textwidth]{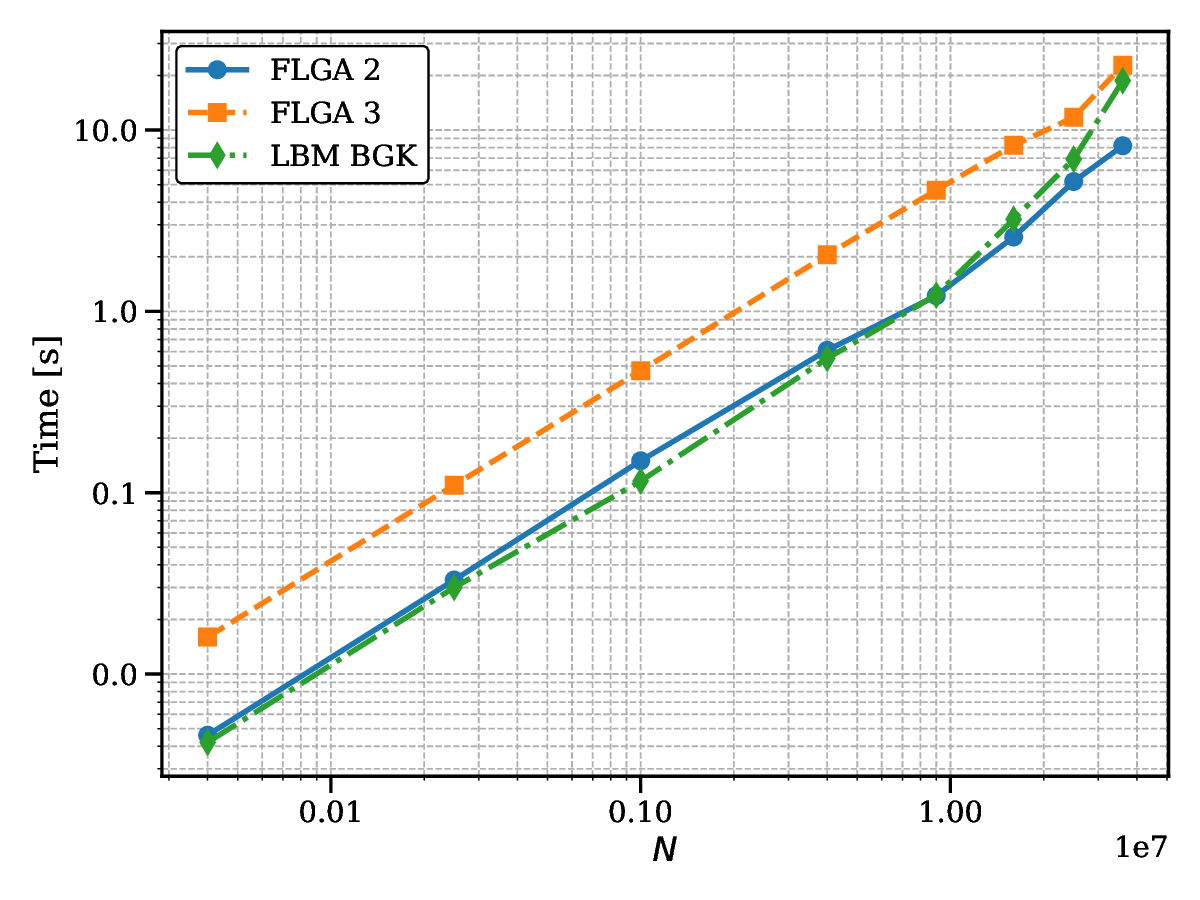}
        
    \caption{Time scaling computing LBM and FLGA for one time-step during the collision operator in the case of a Taylor-Green vortex simulation in Python with M2 processor. For the LBM code, the implementation of Krüger \textit{et al.} (Chapter 13) has been used \cite{kruger2017taylor}. FLGA 2 refers to FLGA with two-body collisions and FLGA 3 with only three-body collisions. }
    \label{fig:time_scaling}
\end{figure}

The effects of the memory overhead are observed in Figure~\ref{fig:time_scaling}, where we can see that LBM using the BGK approximation increases its computational complexity for large lattices, close to FLGA 3 (three-body collisions). Even considering that a more optimal implementation can be used for LBM, the memory allocation and the usage of regular matrices in FLGA (no need for embedding) is an advantage. Additionally, an extra benefit of FLGA is that viscosity depends on the number of equivalence classes and a number of particles to collide, as seen previously. This means that the computational complexity can be adjusted depending on the desired viscosity. For specific simulations, FLGA may represent a great advantage with respect to LBM. According to early benchmarks, by using a lid-driven cavity, FLGA converges faster and with a similar or larger Reynolds numbers compared to LBM when using two, three, and four-body collisions. A more systematic comparison, however, is still required in the future. 

\subsection{Viscosity dependence on equivalence classes, number of particles and collisions}
\label{sec:three_two}
In this section, we will compare how relaxation time and viscosity change with the number of particles undergoing collisions, the number of collisions, and the equivalence classes. 

In ILGA, the relaxation time is reduced when the parameter responsible for the effective number of collisions $C$ and equivalence classes increases. In FLGA, the equilibrium distribution functions are the same and therefore the theoretical derivations are equivalent. This means that for 2D, following \eqref{eq:sigmaxy}, including new equivalence classes is directly proportional with the Reynolds number. On the other hand, increasing the number of times we apply the collision opeartor has no effect in FLGA for high values of $C$. If the number of collisions for each time step is increased, this will be equivalent to increasing the collision rate given by $C$ but with a longer computation time, as we are computing the same operation many times. Equivalently, when increments of $C$ lead to numerical errors, increasing the number of collisions will have the same effect.

\begin{figure}
    
\includegraphics[width=0.48\textwidth]{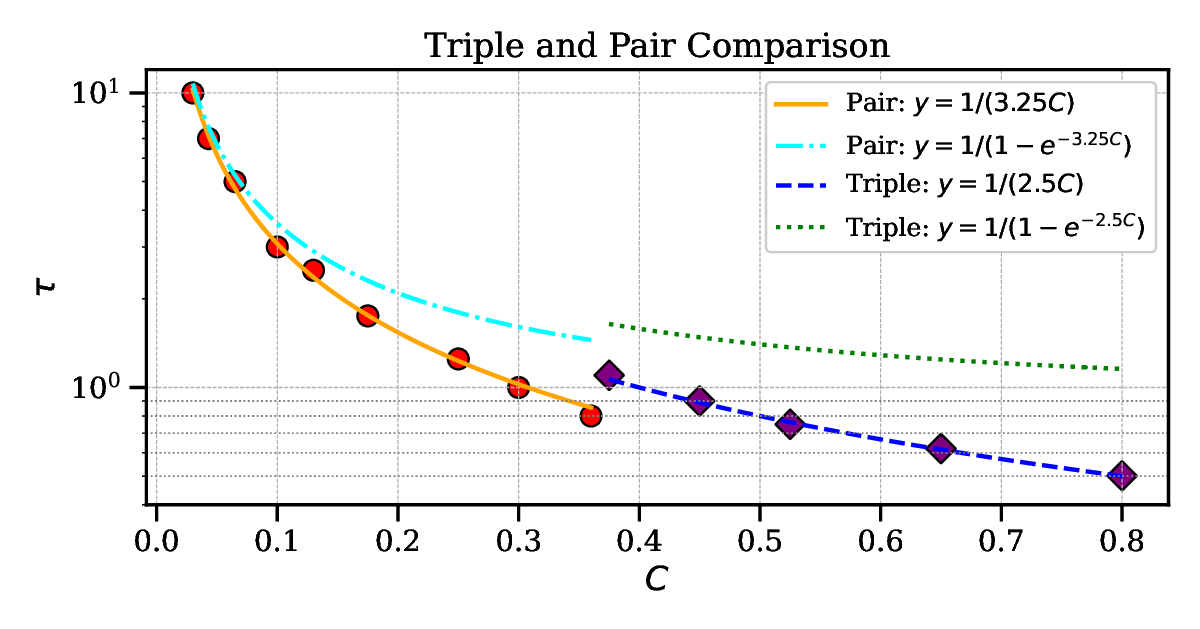}
        
    \caption{Comparison of the relaxation time $\tau$ and the number of collisions $C$ of triple (three-body collisions) and pairs (two-body collisions) in the collision term for a Taylor-Green vortex simulation. The simulation parameters are $L_x=L_y=50$, $u_x=0.1$, $\lambda_i=1$. The exponential curves correspond to the approximation established in ILGA for low deviations of the moments respect the equilibrium. The parameter $\gamma(\lambda)$ next to $C$ has been fitted and $\tau$ has been estimated by comparing with the theoretical equation \eqref{eq:taylor_green}.}
    \label{fig:two_three}
\end{figure}

Increasing the number of particles undergoing the collision (and therefore the number of equivalence classes) is the most effective way to increase the range of the viscosity, being able to decrease it further. However, a theoretical derivation of these is challenging due to the high number of possible equivalence classes and high order of moments in the expressions. Another alternative is to use simulation tests to compute the relaxation time that can be achieved with two-body and three-body collisions. In Fig~\ref{fig:two_three}, we can see the relation of $\tau$ with $C$ in a Taylor-Green vortex simulation, where $\tau$ was estimated heuristically by using the analytical expressions \eqref{eq:taylor_green}. 
As we see, the deviations of the moments respect the equilibrium become larger when we increase $C$, as the exponential approximation becomes worse.
Additionally, increasing the number of equivalence classes with triple collisions increases the difference further. Regarding the relaxation time, triple collisions provide unexpected results for low values of $C$ with accurate results only for $C$ values at least as large as the maximum $C$ for two-body collisions. We also observe a slight jump in the relaxation when transitioning from two to three-body collisions, which is unexpected and whose explanation has not been found. While for two-body collisions the smaller $\tau$ obtained is $0.85$, for three-body collisions, this is reduced to $0.5015$, which is a substantial difference.

\section{Quantum float lattice gas automata}
\label{sec:qflga}
\begin{figure}
	\centering
	\resizebox{0.5\textwidth}{!}{\input{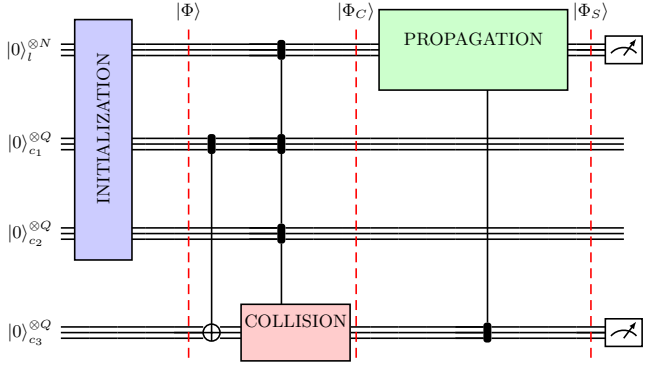}}
    \caption{Quantum circuit scheme for QFLGA. This general diagram uses two-body collisions, but more body collisions can be implemented by adding additional registers $\ket{c_i}$.}
    \label{fig:scheme_qflga}
\end{figure}

Leveraging the simplicity of the model and the ability to compute nonlinear terms directly with the local density functions, we developed a quantum algorithm for FLGA. Following a similar structure as QLGA by Zamora \textit{et al.} \cite{bastida2025efficient}, the Quantum Float Lattice Gas Automata (QFLGA) consists of four steps: Initialization, collision, propagation, and measurement. The algorithm is restricted to one time-step since the difficulties to expand to more time steps are similar to those of Quantum LBM. However, as we will discuss later, this algorithm allows for approximations that could lift off this difficulty. As the propagation and measurement are similar to those of QLBM, we will focus more on the encoding and collision steps. A general diagram of the algorithm can be seen in Fig~\ref{fig:scheme_qflga}.

\begin{itemize}
    \item \textbf{Initialization:} In contrast to LGA, this model uses float numbers as in LBM. This translates to changing the binary or computational basis encoding to an amplitude encoding, where we set the probability to measure a channel $f_i$ at a lattice site $x$ in the amplitude. To simplify the quantum circuit, we have to precompute classically two probabilities appearing in \eqref{eq:collision_f}. First, the density $\rho_l$ at a given lattice site $l$ and second the relative probability of each channel within a lattice. The density was treated as a normalized quantity and therefore it can be easily implemented in the quantum computer, being the probability of having particles in a given site. 

    \begin{equation}
        \rho_l=\frac{\sum_i f_i(l)}{\sum_{x,i} f_i(x)}
    \end{equation}
    
    On the other hand, the probability distributions for each channel $f_i$ will be 
    
    \begin{equation}
    p_j=\frac{f_j(l)}{\sum_i f_i(l)}=\frac{f_j(l)}{\rho_l \sum_{x,i} f_i(x)}=\frac{f_j(l)}{\rho_l}
    \end{equation}

    given that $\sum_{x,i} f_i(x)=1$. Therefore, the initial state can be represented as

    \begin{equation}
        \ket{\phi}=\sum_{l,c}\rho_l p_{c}(l)\ket{l}\ket{c}=\sum_{l,c}f_c(l)\ket{l}\ket{c}
    \end{equation}

   with $\ket{c}$ the channel register and $\ket{l}$ for the lattice sites. In addition to this, we need to encode a second register with the same information. Looking at equation \eqref{eq:collision_f}, we notice that the algorithm relies on the multiplication of the probability of encountering every pair. 
   To solve this, we need two copies of $\phi$ to do the tensor product of both of them with a single lattice register $\ket{l}$. For this, the registers need to be unentangled. With these considerations we can write the state as

    \begin{equation}
        \ket{\Phi}=\sum_{l,c_1,c_2}\sqrt{f_{c_1}(l)f_{c_2}(l)}\ket{l}\ket{c_1}\ket{c_2}
        \label{eq:i_state}
    \end{equation}

\item \textbf{Collision:} The main advantage of using a quantum computer for this algorithm is the ability to collide every lattice site and to obtain the tensor product of all the pair channels at the same time in parallel. 
The main issue with the collision of FLGA is its non-reversibility. In contrast to the classical LGA, where several channels are exchanged by others, in this case we do not add or subtract the same probability from channels $c_1,c_2$ to $c_3,c_4$ as for $c_3,c_4$ to $c_1,c_2$ due to the weights from \eqref{eq:probtrans}. To solve this, we create a third channel register that will act as an ancilla register, entangled with one of the two-channel registers previously created. This way, the computation on the third register will not modify the two previous ones, conserving the information and making the operator reversible and unitary. The main issue of this strategy is the increasing amount of qubits needed to compute several time steps consecutively without measurement. Nonetheless, alleviating this is not in the scope of this work. Another advantage of creating a third register to compute the collision is that we simplify the quadratic dependency of $\rho$ in the denominator from \eqref{eq:i_state}. To do so, we use an ancilla with state $\ket{1}$ when the registers $c_1$ and $c_2$ are in the desired channel. Then, this ancilla can be used as a condition to change the third register $c_3$, obtaining

\begin{equation}
    \frac{f_i^{c+1}}{\rho}= \frac{f_i^{c}}{\rho}+\vartheta_i(jklm)\frac{f_j^c f_k^c}{\rho^2} P_{jk\rightarrow lm}
\end{equation}

However, as $c_3$ is just an entangled copy of one of the above registers, there is a common term of $\rho$ that we can simplify. Therefore, this implementation allows the computation of all the collision pairs with the initialization at no extra cost with the simplifications from FLGA directly in a quantum circuit and without computing macroscopic variables. This common term will appear naturally when we measure only one register, summing over all possible contributions. 


After the collision, the statevector becomes 
{\small

\begin{equation}
    \begin{aligned}
        \ket{\Phi_C} 
        &= I^{\otimes^N} \otimes C^{\otimes^{c}} 
        \left(\sum_{l,c_1,c_2} \sqrt{f_{c_1}(l)f_{c_2}(l)}\ket{l}\ket{c_1}\ket{c_2}\ket{c_1}\right) \\
        &= \sum_{l,c_1,c_2} \sqrt{f_{c_1}(l) f_{c_2}(l)  
        \left(1 - \sum_{c_3,c_4} P_{c_1 c_2\rightarrow c_3 c_4} \right)} 
        \ket{l}\ket{c_1}\ket{c_2}\ket{c_1} \\
        &\quad + \sum_{l,c_1,c_2,c_3 : c_1\neq c_3} 
        \sqrt{f_{c_1}(l) f_{c_2}(l)  
        \sum_{c_4} P_{c_1 c_2\rightarrow c_3 c_4}} 
        \ket{l}\ket{c_1}\ket{c_2}\ket{c_3} \\
        &\quad = \sum_{l,c_1,c_2,c_3}\sqrt{f_{c_1}(l) f_{c_2}(l) P(c_1,c_2,c_3)}\ket{l}\ket{c_1}\ket{c_2}\ket{c_3} 
    \end{aligned}
    \label{eq:i_state}
\end{equation}
}

 splitting the amplitude probability to a different register $c_3$ given by the target channels that undergo a collision with the right probability. $P(c_1,c_2,c_3)$  is used here as a term to generalize the probability in each case.

\item \textbf{Propagation, measurement and post-processing:} 
The propagation step is similar to QLGA, where each channel has its own direction of propagation, following the principles of LGA. To the best knowledge of the authors, the best performing propagation algorithms are the QFT-based propagation \cite{Shakeel_2020} and the parallel-shift propagation \cite{budinski2023efficient}. In this case,  without loss of generality, QFT-based propagation is applied to the third channel register $c_3$. 
After the propagation step, the quantum state becomes
{\small
\begin{equation}
    \begin{aligned}
        \ket{\Phi_S} 
        &= S^{\otimes^{N+c}}\sum_{l,c_1,c_2,c_3}\sqrt{f_{c_1}(l) f_{c_2}(l) P(c_1,c_2,c_3)}\ket{l}\ket{c_1}\ket{c_2}\ket{c_3}\\
        &= \sum_{l,c_1,c_2,c_3}\sqrt{f_{c_1}(l+v_{c_3}) f_{c_2}(l+v_{c_3}) P(c_1,c_2,c_3)}\ket{l}\ket{c_1}\ket{c_2}\ket{c_3} 
    \end{aligned}
    \label{eq:i_state}
\end{equation}
}
with $l+v_{c_3}$ the shift given by the velocity associated to the channel $c_3$. For example for D1Q3 if $c_3$ is a right particle, the final lattice site after the shift is $l+1$, while for a left particle it is $l-1$. The measurement will involve the lattice register $l$ and the third channel $c_3$, obtaining a reduced state where we only measure the selected qubits. Given that we are using amplitude encoding, the noise-mitigation properties of QLGA \cite{bastida2025efficient} will be mostly lost, the algorithm behaving more similarly to QLBM in this regard \cite{doi:10.1142/S0219749921500398}. As a result of the repeated partial measurement of the quantum state, we obtain the following probabilities for the cases when $c_1=c_3$ and when $c_1\neq c_3$

{\small
\begin{equation}
\begin{aligned}
&P(c_1=c_3,l)=f_{c_1}(l+v_{c_1})\left(1-\sum_{c_2,c_3,c_4} f_{c_2}(l+v_{c_1}) P_{c_1 c_2\rightarrow c_3 c_4}\right) \\
&P(c_1\neq c_3,l)=\sum_{c_1,c_2,c_4} f_{c_1}(l+v_{c_3})f_{c_2}(l+v_{c_3}) P_{c_1 c_2\rightarrow c_3 c_4}
\end{aligned}
\end{equation}
}
which means that $f_c^{(1)}(l)$ is 
{\small
\begin{equation}
f_c^{(1)}(l)=f_{c}(l+v_{c})+\sum_{i,j,k} \vartheta(i,j,k,c)f_i(l+v_{c}) f_j(l+v_{c})P_{ij\rightarrow kc} 
\end{equation}
}

The post-processing of the algorithm will change depending on the measurement strategy or if we use an emulator or an actual quantum device. The main step will involve the calculation of the local density $\rho$ and the density probabilities $f_i$. This is a significant advantage with respect to QLBM, reducing the classical overhead with nonlinear terms.

\item \textbf{Simulating multiple-time steps}: Simulating multiple time-steps with this algorithm is restricted in a similar way as in QLGA. The problem here does not stem from the specific quantum circuit utilized but the encoding and type of solver. Proved already by Schalkers \textit{et al.} \cite{Schalkers2024}, neither the computational basis encoding nor amplitude encoding allow for concatenating collision and propagation operators without measurement.
While a concatenation is possible by encoding every channel in every lattice site, this is inefficient if using QLGA or QFLGA algorithms and does not provide any advantage. Even so, different strategies to accelerate the quantum algorithm and limit the number of measurements are possible. One possible strategy could be to accelerate the simulation by approximating the dynamics by skipping several collision steps, allowing for quantum circuits for multiple time-steps including a more extensive propagation. Another alternative is to include an approximation for $f_k^c$ in \eqref{eq:collision_f}, to estimate the colliding state. This is clear when we think of the structure of the circuit. There are two parts, first the state to modify stored in the register $c_3$ copied from $c_1$, which is equivalent to $f_i^c$ and the state it collides with $f_k^c$ given by $c_2$. This allows us to initialize the neighbour channels in each lattice site in a fourth or fifth register $c_4$ or $c_5$ to be used later on as a proxy of $f_k^c$. These states could also collide with previous initialized registers to obtain a closer approximation.

\end{itemize}

\section{Quantum results and challenges}


First, we focused on a simple nonlinear case to test the equivalence between FLGA and its quantum version (QFLGA) presented in Section~\ref{sec:qflga}. Therefore, a shockwave 1D case was targeted as an initial showcase.

\begin{figure}[H]
	\centering
    \scalebox{0.5}{\input{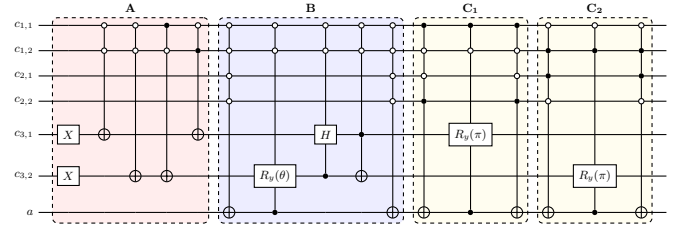}}
    \caption{Quantum circuit from collision operator of QFLGA algorithm }
    \label{fig:collision-qflga}
\end{figure}

An example of the implementation of the collision operator for D1Q3 can be seen in Figure~\ref{fig:collision-qflga}. The red box \textbf{A} contains the entanglement of the first to third channel register. Then the blue box \textbf{B} targets the state $\ket{00}$ of each channel register $c_1$ and $c_2$ corresponding with the rest particle. When both states are in rest, an $R_y$ gate with angle $\theta$ is applied, followed by a Hadamard gate to split the state to right and left channels. This angle depends of the transition probability $P_{ij\rightarrow kl}$ given by \eqref{eq:probtrans}. Finally, the yellow boxes \textbf{$C_1$} and \textbf{$C_2$} are in charge of the second collision, transforming the state $\ket{\rightarrow \leftarrow}$ (right and left) to $\ket{\cdot \cdot}$ (rest and rest). One limitation of this implementation is that the pair $\ket{\rightarrow \leftarrow}$ is coupled to the right channel of the register $c_3$, while the pair $\ket{\leftarrow \rightarrow}$ is to the left channel. This results in probability being subtracted from only the right channel or only the left channel, instead of from both, in each respective case. 
In practice, this means that the maximum probability to split is smaller, which limits the parameter $\lambda$ to half. 
This affects for example to the ability to have overrelaxation in the quantum version. However, ideas such as implementing void particles to renormalize later on the probabilities or implementing a slightly different initialization could solve this issue. 

\begin{figure}
    \centering
    \includegraphics[width=0.48\textwidth]{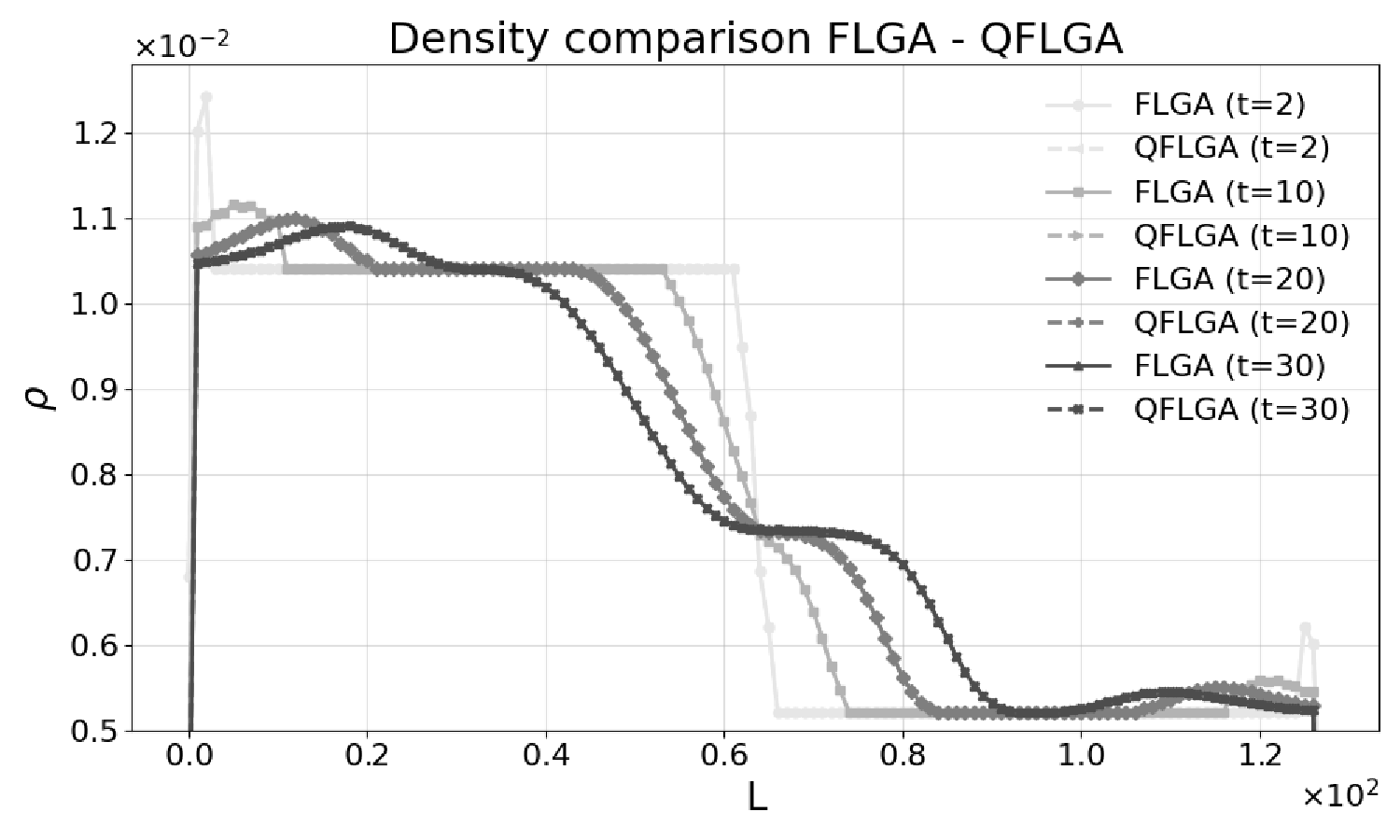}

    \caption{Comparison between QFLGA and FLGA for a shockwave in 1D using $L=128$ and $C=0.5$} 
    \label{fig:shockwave_qflga}
\end{figure}

Figure~\ref{fig:shockwave_qflga} shows the results from the Qiskit \cite{qiskit2024} simulation using the described collision circuit.  The agreement between FLGA and QFLGA results confirms the validity of the QFLGA method.  For validation purposes, a simplified boundary condition was used: at each boundary, amplitudes of incoming channels were added to those of outgoing channels. 
In LBM, fixed values at the boundary are usually used instead of adding to the previous ones, not conserving mass.  Nevertheless, the same results can be obtained for other boundary conditions without loss of generality.

The expansion of the method for D2Q9 to model Navier-Stokes equations is possible, following the scheme presented in Fig~\ref{fig:scheme_qflga}. However, the number of qubits to encode the three collision operator (9) with the number of qubits for the lattice (for a small lattice of 32x32 sites, we require 10 qubits) and the ancilla for boundary conditions (2-3) is too large to demonstrate results for non-linear flows. Therefore, this expansion is left as future research. The implementation of the collision operator for this method is equivalent to the one for D1Q3 presented in Figure~\ref{fig:collision-qflga}. Specifically, nine equivalence classes are needed with 2-8 collisions in each of them, increasing the number of operations and the length of the circuit significantly, making it slow to emulate in classical devices. 

\section{Conclusions}


In this article, we presented a novel algorithm for fluid dynamics as an intermediary step between Lattice Gas Automata (LGA) and Lattice Boltzmann Method (LBM). 
Following the initial method developed by Blommel \textit{et al.} on Integer Lattice Gas Automata (ILGA) \cite{PhysRevE.97.023310}, we expanded it to a mesoscopic scale by using an ensemble average with float numbers. We have called this new variant Float Lattice Gas Automata (FLGA). While the main analytical results and theory are the same as for its microscopic counterpart, this model allows for faster simulations, where only one collision operator is needed. Further simplifications allow for over-relaxation with the relaxation time $\tau<1$ and a smaller number of terms in the collision operator. This model establishes a connection between Molecular Dynamics simulations (MD) and LBM by allowing the calculation of the probability distribution of particles without a direct simulation of them at a mesoscopic level. The interaction between particles can be modified with $\lambda$  (effective collision rate) and the transition probability $P_{ij\rightarrow kl}$

In contrast to \cite{PhysRevE.97.023310}, the model was tested for different fluid flows such as shockwaves in 1D and Taylor-Green vortex and lid-driven cavity in 2D, obtaining accurate results when compared with LBM or the analytical solution. We also studied the relation between $\tau$ and $C$, a parameter responsible for the effective collision of the system. Lattice gas automata methods present difficulties in finding this value, as it is generated by the collisions and not given as an input in the simulation. However, we found that it is generally possible to estimate $\tau$ preceding the simulation following a heuristic approach. Nonetheless, several details, such as the most efficient $\lambda$ parameters to be used (effective rate of each collision equivalence class) or the effect of boundary conditions in the relation of $\tau$ and $C$ are left as open questions for future research. 

The expansion to three-body collisions and the computational complexity of the method compared with LBM were also explored. Regarding the latter, we found a similar computational complexity with a slightly smaller use of data that, for a large number of lattice sites, becomes relevant. In specific, data access becomes easier, as everything is directly computed using the local distribution functions $f_i$ with no calculation of macroscopic velocities and densities. 
 The ability to change the computational complexity by increasing or decreasing collisions also allows more freedom to control it depending on the target viscosity. However, we note that our comparison was carried out using Python and a more detailed analysis between FLGA and LBM is still required.
 
As an extension to solve fluid flows with high Reynolds numbers, requiring a large number of lattice sites, we developed a quantum version of the algorithm that holds such potential. Quantum float lattice gas automata were tested for a shockwave D1Q3 case, obtaining the same results as its classical counterpart. While the quantum algorithm is restricted to one-time step, requiring measurement and state encoding to continue the simulation, early attempts suggest that the method could be promising to avoid this issue in the future. In specific, colliding approximated local functions $\tilde{f_i}$ with the correct local functions $f_i$, or using a Carleman linearization approach, taking advantage of the reduced number of terms of the model, could alleviate this problem. The strength of this algorithm is its simplicity, allowing for computing non-linear terms with a relatively small number of quantum gates. 

\section{Acknowledgments} 
We would like to express our sincere gratitude to Niccolò Fonio, doctoral student from Aix-Marseille Université for insightful discussions regarding the possibilities for a quantum algorithm for several time steps and the classical understanding of the model.

\bibliography{apssamp}

\end{document}